\documentclass[copyright]{eptcs}

\providecommand{\event}{MARS 2018}

\usepackage{underscore}
\usepackage{url}
\usepackage{breakurl}

\usepackage{listings}
\usepackage{cadp-lotos}
\usepackage{cadp-lnt}

\usepackage{xcolor}

\definecolor{grey}{cmyk}{0,0,0,0.8}
\definecolor{gray}{cmyk}{0,0,0,0.5}
\hypersetup{bookmarks=true,bookmarksopen=false,colorlinks=true,linkcolor=grey,citecolor=grey,urlcolor=grey}

\newcommand{\B}[1]{\mbox{\bf #1}}
\newcommand{\I}[1]{\mbox{\em #1\/}}

\newcommand{\T}[1]{\mbox{\tt #1}}

\newcommand{\implies}{\Rightarrow}
\newcommand{\CAESARADT}{C{\AE}SAR.ADT}

\newcommand{\SPEC}[1]{\colorbox{gray}{\color{white}#1}}

\newcommand{\IGNORE}[1]{}

\title{Comparative Study of Eight Formal Specifications \\ of the Message Authenticator Algorithm}

\def\titlerunning{Comparative Study of Eight Formal Specifications of the Message Authenticator Algorithm}

\author{Hubert Garavel \quad\qquad Lina Marsso
   \institute{INRIA Grenoble, France}
   \institute{Univ. Grenoble Alpes, LIG, F-38000 Grenoble, France}
   \institute{CNRS, LIG, F-38000 Grenoble, France}
   \email{Hubert.Garavel@inria.fr \quad\qquad Lina.Marsso@inria.fr}
}

\def\authorrunning{H.~Garavel \& L.~Marsso}

\begin{document}

\maketitle

\begin{abstract}
	The Message Authenticator Algorithm (MAA) is one of the first cryptographic functions for computing a Message Authentication Code. Between 1987 and 2001, the MAA was adopted in international standards (ISO 8730 and ISO 8731-2) to ensure the authenticity and integrity of banking transactions. In 1990 and 1991, three formal, yet non-executable, specifications of the MAA (in VDM, Z, and LOTOS) were developed at NPL. Since then, five formal executable specifications of the MAA (in LOTOS, LNT, and term rewrite systems) have been designed at INRIA Grenoble. This article provides an overview of the MAA and compares its formal specifications with respect to common-sense criteria, such as conciseness, readability, and efficiency of code generation.
\end{abstract}

\section{Introduction}

	To handle real problems, formal methods should be capable of describing the different facets of a system: data structures, sequential algorithms, concurrency, real time, probabilistic and stochastic aspects, hybrid systems, etc.
	In the present article, we address the two former points. In most case studies, the data structures and their algorithms are relatively simple, the most complex ones being trees, which are explored using breadth-first or depth-first traversals, etc. Contrary to such commonplace examples, cryptographic functions exhibit more diverse behaviour, as they rather seek to perform irregular computations than linear ones.

	To explore this dimension, we consider the {\em Message Authenticator Algorithm\/} (MAA, for short), a pioneering cryptographic function designed in the mid-80s at the National Physical Laboratory (NPL, United Kingdom). The MAA was adopted in two international standards (ISO 8730 and ISO 8731-2) and served, between 1987 and 2001, to secure the authenticity and integrity of banking transactions. The MAA also played a role in the history of formal methods, as the NPL developed, in the early 90s, three formal specifications of the MAA in VDM, Z, and LOTOS abstract data types.

	The present article revives these early efforts by examining, twenty-five years later, how the new generation of formal methods can cope with the MAA case study. The article is organized as follows. Section~\ref{OVERVIEW} presents the MAA from both an historical and technical perspective. Section~\ref{MODELS} introduces the eight formal specifications of the MAA we are aware of. Section~\ref{DISCUSSION} discusses some key modelling issues that arise when specifying the MAA. Section~\ref{VALIDATION} precises how the formal specifications have been validated and which issues have been uncovered. Section~\ref{CONCLUSION} gives concluding remarks. Annexes~\ref{ERRATA-8730} and ~\ref{ERRATA-8731} report errors found in the MAA test vectors prescribed by ISO standards 8730 and 8731-2. Finally, Annexes~\ref{ANNEX-LOTOS} and \ref{ANNEX-LNT} provide two formal specifications of the MAA in LOTOS and LNT, which are novel contributions.

\section{The Message Authenticator Algorithm (MAA)}
\label{OVERVIEW}

	In data security, a {\em Message Authentication Code\/} (MAC) is a short sequence of bits that is computed from a given message; the MAC ensures both the authenticity and integrity of the message, i.e., that the message sender is the stated one and that the message contents have not been altered.
	A MAC is more than a mere checksum, as it must be secure enough to defeat attacks; its design usually involves cryptographic keys shared between the message sender and receiver.
	One of the first MAC algorithms to gain widespread acceptance was the MAA, which we now present in more detail.

\subsection{History of the MAA}

	The MAA was designed in 1983 by Donald Watt Davies and David Clayden at NPL, in response to a request of the UK Bankers Automated Clearing Services \cite{Davies-Clayden-83} \cite{Davies-85}. Its authors were formerly involved in the detailed design and development of Pilot ACE ({\em Automatic Computing Engine\/}), an early computer based on original designs of Alan Turing. Donald Watt Davies (1924--2000) is a founding father of computer science, also well known for his pioneering work on computer networks and packet switching in the mid-60s\footnote{Biographic information about D. W. Davies can be found from \url{http://en.wikipedia.org/wiki/Donald_Davies} and \url{http://thelinuxmaniac.users.sourceforge.net/docs/be/chc61}.}.
	Shortly after its design, the MAA became standardized at the international level in two complementary ISO banking standards:

\begin{itemize}
	\item The ISO international standard 8730 (published in 1986 \cite{ISO-8730:1986} and revised in 1990 \cite{ISO-8730:1990}) specifies methods and procedures for protecting messages exchanged between financial institutions. Such a protection is based on secret keys symmetrically shared between these institutions and on the computation of a MAC for each message exchanged.

	The 1986 version of this standard \cite{ISO-8730:1986} was independent from any particular algorithm for MAC computation. Such independence was slightly undermined by the 1990 revision of this standard \cite{ISO-8730:1990}, which added two annexes~D and~E providing test vectors (i.e., MAC values for a few sample messages and given keys) computed using two specific algorithms (DEA and MAA) presented hereafter. A technical corrigendum was later issued in 1999 \cite{ISO-8730:1999} to address the Year-2000 problem, without any impact of the MAC computation itself.

	\item The ISO international standard 8731 has two distinct parts, each devoted to an approved algorithm for MAC computation that can be used in the security framework specified by ISO~8730. Both algorithms are mutually exclusive, in the sense that using only one of them is deemed to be sufficient for authenticating messages:

\begin{itemize}
	\item Part 1 (i.e., ISO 8731-1) describes the DEA ({\em Data Encryption Algorithm\/}) which is a CBC-MAC ({\em Cipher Block Chaining Message Authentication Code\/}) based on the DES standard cryptographic algorithm. The DEA is not addressed in the present article.

	\item Part 2 (i.e., ISO 8731-2, published in 1987 \cite{ISO-8731-2:1987} and slightly revised in 1992 \cite{ISO-8731-2:1992}) describes the MAA itself. An equivalent, freely available specification of the MAA can also be found in a 1988 NPL technical report written by the designers of the MAA \cite{Davies-Clayden-88}.
\end{itemize}
\end{itemize}

	Later, cryptanalysis of MAA revealed several weaknesses, including feasible brute-force attacks, existence of collision clusters, and key-recovery techniques \cite{Preneel-vanOorschot-95} \cite{Preneel-vanOorschot-96} \cite{Rijmen-Preneel-DeWin-96} \cite{Preneel-97} \cite{Preneel-Rumen-vanOorschot-97} \cite{Preneel-vanOorschot-99}. After such discoveries, MAA ceased to be considered as secure enough and was withdrawn from ISO standards in 2002 \cite{Preneel-11}.

\subsection{Overview of the MAA}

	Nowadays, Message Authentication Codes are computed using different families of algorithms based on either cryptographic hash functions (HMAC), universal hash functions (UMAC), or block ciphers (CMAC, OMAC, PMAC, etc.). Contrary to these modern approaches, the MAA was designed as a standalone algorithm that does not rely on any preexisting hash function or cipher.

	In this section, we briefly explain the principles of the MAA. More detailed explanations can be found in \cite{Davies-85}, \cite{Davies-Clayden-88} and \cite[Algorithm 9.68]{Menezes-vanOorschot-Vanstone-96}.

	The MAA was intended to be implemented in software and to run on 32-bit computers. Hence, its design intensively relies on 32-bit words (called {\em blocks\/}) and 32-bit machine operations.

	The MAA takes as inputs a key and a message. The key has 64 bits and is split into two blocks $J$ and $K$. The message is seen as a sequence of blocks. If the number of bytes of the message is not a multiple of four, extra null bytes are added at the end of the message to complete the last block. The size of the message should be less than 1,000,000 blocks; otherwise, the MAA result is said to be undefined; we believe that this restriction, which is not inherent to the algorithm itself, was added in the ISO 8731-2 standard to provide MAA implementations with an upper bound (four megabytes) on the size of memory buffers used to store messages.

	The MAA produces as output a block, which is the MAC value computed from the key and the message. The fact that this result has only 32 bits proved to be a major weakness enabling cryptographic attacks; MAC values computed by modern algorithms now have a much larger number of bits. Apart from the aforementioned restriction on the size of messages, the MAA behaves as a totally-defined function; its result is deterministic in the sense that, given a key and a message, there is only a single MAC result, which neither depends on implementation choices nor on hidden inputs, such as nonces or randomly-generated numbers.

	The MAA calculations rely upon conventional 32-bit logical and arithmetic operations, among which: \T{AND} (conjunction), \T{OR} (disjunction), \T{XOR} (exclusive disjunction), \T{CYC} (circular rotation by one bit to the left), \T{ADD} (addition), \T{CAR} (carry bit generated by 32-bit addition), \T{MUL} (multiplication, sometimes decomposed into \T{HIGH\_MUL} and \T{LOW\_MUL}, which denote the most- and least-significant blocks in the 64-bit product of a 32-bit multiplication).
	On this basis, more involved operations are defined, among which \T{MUL1} (result of a 32-bit multiplication modulo $2^{32}-1$), \T{MUL2} (result of a 32-bit multiplication modulo $2^{32}-2$), \T{MUL2A} (faster version of \T{MUL2}), \T{FIX1} and \T{FIX2} (two unary functions\footnote{The names \T{FIX1} and \T{FIX2} are borrowed from \cite[pages 36 and 77]{Munster-91-a}.} respectively defined as $x \rightarrow \T{AND} (\T{OR} (x, \T{A}), \T{C})$ and $x \rightarrow \T{AND} (\T{OR} (x, \T{B}), \T{D})$, where \T{A}, \T{B}, \T{C}, and \T{D} are four hexadecimal block constants \T{A} = 02040801, \T{B} = 00804021, \T{C} = BFEF7FDF, and \T{D} = 7DFEFBFF).
	The MAA operates in three successive phases:

\begin{itemize}
	\item The {\em prelude\/} takes the two blocks $J$ and $K$ of the key and converts them into six blocks $X_0$, $Y_0$, $V_0$, $W$, $S$, and $T$. This phase is executed once. After the prelude, $J$ and $K$ are no longer used.

	\item The {\em main loop\/} successively iterates on each block of the message. This phase maintains three variables $X$, $Y$, and $V$ (initialized to $X_0$, $Y_0$, and $V_0$, respectively), which are modified at each iteration. The main loop also uses the value of $W$, but neither $S$ nor $T$.

	\item The {\em coda\/} adds the blocks $S$ and $T$ at the end of the message and performs two more iterations on these blocks. After the last iteration, the MAA result, noted $Z$, is $\T{XOR} (X, Y)$.
\end{itemize}

	In 1987, the ISO 8731-2 standard \cite[Sect.~5]{ISO-8731-2:1987} introduced an additional feature (called {\em mode of operation\/}), which concerns messages longer than 256 blocks (i.e., 1024 bytes) and which, seemingly, was not present in the early MAA versions designed at NPL. Each message longer than 256 blocks must be split into {\em segments\/} of 256 blocks each, with the last segment possibly containing less than 256 blocks.
	The above MAA algorithm (prelude, main loop, and coda) is applied to the first segment, resulting in a value noted $Z_1$. This block $Z_1$ is then inserted before the first block of the second segment, leading to a 257-block message to which the MAA algorithm is applied, resulting in a value noted $Z_2$. This is done repeatedly for all the $n$ segments, the MAA result $Z_i$ computed for the $i$-th segment being inserted before the first block of the $(i+1)$-th segment. Finally, the MAC for the entire message is the MAA result $Z_n$ computed for the last segment.

\subsection{Informal Specifications of the MAA}

	We consider the 1988 NPL technical report \cite{Davies-Clayden-88} to be the reference document for the MAA definition in natural language. Indeed, this technical report is freely available from the NPL library or can be downloaded from the web, whereas the (withdrawn) ISO standards 8730 and 8731-2 need to be purchased. The algorithm described in \cite{Davies-Clayden-88} is identical to the MAA definition given in ISO~8731-2.

	Moreover, \cite{Davies-Clayden-88} provides the source code of two different implementations of the MAA, in BASIC (227 lines\footnote{In the present paper, when counting lines of code, we exclude blank lines, comments, as well as predefined libraries. Concerning the MAA implementation in BASIC, we also exclude all \T{PRINT} statements, mostly intended for debugging purpose.}) and C (182 lines\footnote{We exclude all \T{prinf} statements, as well as five non-essential functions (\T{menu1}, \T{inmain}, \T{mainloop1}, \T{fracsec}, and \T{guesstime}), only retaining case 5 in the \T{main} function, and unfolding multiple instructions present on the same lines.}).
	None of these implementations supports the aforementioned ``mode of operation''; we therefore added 31~lines of C~code implementing this missing functionality.
	Although the C~code was written in 1987 for the Turbo~C and Zorland compilers, it still compiles and executes properly today after a few simple corrections, provided that long integers are set to 32~bits\footnote{For instance, using GCC with options \T{-m32 -std=c90}.}.

\subsection{Test Vectors for the MAA}

	There are two official sources of test vectors for the MAA:

\begin{itemize}
	\item \cite[Sections~15 to 20]{Davies-Clayden-88} provides a series of tests vectors contained in six tables, which can also be found in \cite[Annex~A]{ISO-8731-2:1992}. These test vectors specify, for a few given keys and messages, the expected values of intermediate calculations (e.g., \T{MUL1}, \T{MUL2}, \T{MUL2A}, prelude, main loop, etc.) and the expected MAA results for the entire algorithm. The ``mode of operation'' is not tested as the messages considered contain either 2 or 20 blocks, i.e., less than 256 blocks.

	\item Another series of test vectors that take into account the ``mode of operation'' can be found in \cite[Annex E]{ISO-8730:1990}. More precisely, Annex~E.3.3 gives expected values for an execution of the prelude, Annex~E.3.4 gives results for an 84-block message, and Annex~E.4 gives results for a 588-block message.
\end{itemize}

\noindent In both series of test vectors, we found mistakes, which we document and for which we give corrections in Annexes~\ref{ERRATA-8730} and \ref{ERRATA-8731} of the present article.

\section{Formal Specifications of the MAA}
\label{MODELS}

	As far as we know, not less than eight formal specifications have been produced for the MAA. We present each of them in turn, drawing a clear distinction between non-executable and executable specifications. To unambiguously designate these specifications, we adopt the following naming convention: \SPEC{LANG-XX} refers to the formal specification written in language LANG during year XX.

\subsection{Non-Executable Formal Specifications}

	For cryptographic protocols, an informal specification is often not precise enough, and the MAA makes no exception. For instance, G. I. Parkin and G. O'Neill devoted four pages in \cite[Sect.~3]{Parkin-ONeill-90} and \cite[Sect.~3]{Parkin-ONeill-91} to discuss all possible interpretations of the MAA definition in natural language. The need for unambiguous specifications was certainly felt by stakeholders, as three formal specifications of the MAA were developed at NPL in the early 90s, as part of a comparative study in which common examples were modelled using various formal methods. All these specifications were {\em non-executable}, in the sense that MAA implementations had to be developed manually and could not be automatically derived from the formal specifications --- at least, using the software tools available at that time. Let us briefly review these specifications:

\begin{itemize}
	\item \SPEC{VDM-90}\,: In 1990, G. I. Parkin and G. O'Neill designed a formal specification of the MAA in VDM \cite{Parkin-ONeill-90} \cite{Parkin-ONeill-91}. To our knowledge, their work was the first attempt at applying formal methods to the MAA. This attempt was clearly successful, as the VDM specification became a (non-authoritative) annex in the 1992 revision of the ISO standard defining the MAA \cite[Annex~B]{ISO-8731-2:1992}.
	This annex is concise (9 pages, 275 lines) and its style is close to functional programming. Due to the lack of VDM tools, its correctness could only be checked by human proofreading.
	Three implementations in C \cite[Annex~C]{Parkin-ONeill-90}, Miranda \cite[Annex~B]{Parkin-ONeill-90}, and Modula-2 \cite{Lampard-91} were written by hand along the lines of this VDM specification.

	\item \SPEC{Z-91}\,: In 1991, M. K. F. Lai formally specified the MAA using the set-theoretic Z notation. Based upon Knuth's ``literate programming'' approach, this work resulted in a 57-page technical report \cite{Lai-91}, in which formal fragments (totalling 608 lines of Z~code) are inserted in the natural-language description of the MAA. This formal specification was designed to be as abstract as possible, not to constrain implementations unduly, and it was lightly verified using a type-checking tool.

	\item \SPEC{LOTOS-91}\,: In 1991, Harold~B. Munster produced another formal specification of the MAA in LOTOS presented in a 94-page technical report \cite{Munster-91-a}\footnote{This report and its LOTOS specification are available on-line from \url{ftp://ftp.inrialpes.fr/pub/vasy/publications/others/Munster-91-a.pdf} and \url{ftp://ftp.inrialpes.fr/pub/vasy/demos/demo_12/LOTOS/maa_original.lotos}, respectively.}.
	This specification (16 pages, 438 lines) uses only the data part of LOTOS (namely, abstract data types inspired from the ACT~ONE language \cite{Ehrig-Mahr-85} \cite{DeMeer-Roth-Vuong-92}), importing the predefined LOTOS libraries for octets, strings, natural numbers in unary, binary, and decimal notation; the behavioural part of LOTOS, which serves to describe concurrent processes, is not used at all.
	 This specification is mostly declarative, and not directly executable, for at least two reasons:
\begin{itemize}
	\item Many computations are specified using the predefined type \T{Nat} that defines natural numbers in unary notation, i.e., numbers built using the two Peano constructor operations $\T{0} : \rightarrow \T{Nat}$ and $\T{succ} : \T{Nat} \rightarrow \T{Nat}$. On this basis, the usual arithmetic operations (addition, multiplication, etc.) are defined equationally. In practice, such a simple encoding for \T{Nat} cannot feasibly implement the large 32-bit numbers manipulated in MAA calculations.

	\item The full expressiveness of LOTOS equations is used in an unconstrained manner, making it necessary at many places to invert non-trivial user-defined functions. For instance, given a conditional equation of the form $x = g(y) \implies f (x) = y$, evaluating $f(z)$ requires to compute $g^{-1}(z)$.
	Such situations arise, in a more involved way, with $f = \T{NAT}$ and $g = \T{NatNum}$, $f = \T{MUL1}$ and $g = \T{NatNum}$, $f = \T{PAT}$ and $g = \T{BitString}$, $f = \T{BYT}$ and $g = \T{BitString}$, $f = \T{MAC}$ and $g = \T{Flatten}$, etc.
\end{itemize}

	Interestingly, such executability issues are not discussed in \cite{Munster-91-a}. Instead, the report stresses the intrinsic difficulty of describing partial or incompletely-specified functions in LOTOS, the equational semantics of which requires functions to be totally defined. Such difficulty is stated to be a major limitation of LOTOS compared to VDM and Z, although the report claims that LOTOS is clearly superior to these methods as far as the description of communication protocols is concerned.
\end{itemize}

\subsection{Executable Formal Specifications}
\label{REC-TARGETS}

	As a continuation of the work undertaken at NPL, five formal specifications of the MAA have been developed at INRIA Grenoble. These specifications are {\em executable}, in the sense that all expressions that contain neither free variables nor infinite recursion can be given to some interpretation engine and evaluated to produce relevant results. But {\em executable\/} also means that these specifications can be compiled automatically (e.g., using the translators of the CADP toolbox \cite{Garavel-Lang-Mateescu-Serwe-13}) into some executable program that will be run to generate the expected results. Let us review these five specifications:

\begin{itemize}
	\item \SPEC{LOTOS-92}\,: In 1992, Hubert Garavel and Philippe Turlier, taking \SPEC{LOTOS-91} as a starting point, gradually transformed it to obtain an executable specification from which the \CAESARADT\ compiler \cite{Garavel-89-c} \cite{Garavel-Turlier-93} could generate C~code automatically. Their goal was to remain as close as possible to the original LOTOS specification of Harold B. Munster, so as to demonstrate that limited changes were sufficient to turn a non-executable LOTOS specification into an executable one.
	The aforementioned causes of non-executability in \SPEC{LOTOS-91} were addressed by fulfilling the additional semantic constraints set on LOTOS by the \CAESARADT\ compiler to make sure that LOTOS specifications are executable:

\begin{itemize}
	\item The algebraic equations, which are not oriented in standard LOTOS, were turned into term rewrite rules, which are oriented from left to right and, thus, more amenable to efficient translation.

	\item A distinction was made between constructor and non-constructor operations, and the discipline of ``free'' constructors required by \CAESARADT\ \cite{Garavel-89-c} was enforced: namely, each rule defining a non-constructor $f$ must have the form either ``$f (p_1, ..., p_n) \rightarrow e$\/'' or ``$c_1, ..., c_m \implies f (p_1, ..., p_n) \rightarrow e$\/'', where each $p_i$ is a term containing only constructors and free variables, and where $c_1$, ..., $c_m$, and $e$ are terms whose variables must be also present in some $p_i$.

	\item To avoid issues with the unary notation of natural numbers, the \T{Nat} sort was implemented manually as a C~type (32-bit unsigned integer). Similarly, a few operations on sort \T{Nat} (integer constants, addition, multiplication, etc.) were also implemented by manually written C~functions --- the ability to import externally defined C types and functions, and to combine them with automatically generated C~code being a distinctive feature of the \CAESARADT\ compiler. Additionally, all occurrences of the sort \T{BitString} used for the binary notation of natural numbers, octets, and blocks were eliminated from the MAA specification.
\end{itemize}

	This resulted in a 641-line LOTOS specification, together with two C~files (63 lines in total) implementing the LOTOS sorts and operations defined externally. The \CAESARADT\ compiler translated this LOTOS specification into C~code that, combined with a small handwritten main program (161 lines of C~code), could compute the MAC value corresponding to a message and a key.

	\item \SPEC{LNT-16}\,: In February 2016, Wendelin Serwe manually translated \SPEC{LOTOS-92} into LNT \cite{Champelovier-Clerc-Garavel-et-al-10-v6.7}, which is the most recent specification language supported by the CADP toolbox and the state-of-the-art replacement for LOTOS \cite{Garavel-Lang-Serwe-17}. This translation was done in a systematic way, the goal being to emphasize common structure and similarities between the LOTOS and LNT specifications. The resulting 543-line LNT specification thus has the style of algebraic specifications and functional programs, relying massively on pattern matching and recursive functions. The handwritten C code imported by the LOTOS specification was reused, almost as is, for the LNT specification.

	\item \SPEC{REC-17}\,: Between September 2016 and February 2017, Hubert Garavel and Lina Marsso undertook the translation of \SPEC{LOTOS-92} into a term rewrite system\footnote{Actually, it is a conditional term rewrite system with only six conditional rewrite rules that, if needed, can easily be turned into non-conditional rewrite rules as explained in \cite{Garavel-Marsso-17}.}. This system was encoded in the simple language REC proposed in \cite[Sect.~3]{Duran-Roldan-Balland-et-al-09} and \cite[Sect.~3.1]{Duran-Roldan-Bach-10}, which was lightly enhanced to distinguish between free constructors and non-constructors.

	Contrary to higher-level languages such as LOTOS or LNT, REC is a purely theoretical language that does not allow to import external fragments of code written in a programming language. Thus, all types (starting by the most basic ones, such as \T{Bit} and \T{Bool}) and their associated operations were exhaustively defined ``from scratch'' in the REC language.
	To address the aforementioned problem with natural numbers, two different types were defined: a \T{Nat} used for ``small'' counters, the values of which do not exceed a few thousands, and a \T{Block} type that represents the 32-bit machine words used for MAA calculations. The \T{Nat} was defined in the Peano-style unary notation, while the \T{Block} sort was defined in binary notation (as a tuple sort containing or four octets, each composed of eight bits). To provide executable definitions for the modular arithmetic operations on type \T{Block}, the REC specification was equipped with 8-bit, 16-bit, and 32-bit adders and multipliers, somehow inspired from the theory of digital circuits. To check whether the MAA calculations are correct or not, the REC specification was enriched with 203~test vectors \cite[Annexes B.18 to B.21]{Garavel-Marsso-17} originating from diverse sources.

	The resulting REC specification has 1575 lines and contains 13~sorts, 18~constructors, 644~non-constructors, and 684~rewrite rules.  It is minimal, in the sense that each sort, constructor, and non-constructor is actually used (i.e., the specification does not contain ``dead'' code). As far as we are aware, it is one of the largest handwritten term rewrite systems publicly available. Parts of this specification (e.g., the binary adders and multipliers) are certainly reusable for other purposes. However, it is fair to mention that term rewrite systems are a low-level theoretical model that does not scale well to large problems, and that it took considerable effort to come up with a REC specification that is readable and properly structured.

	Using a collection of translators\footnote{\url{http://gforge.inria.fr/scm/viewvc.php/rec/2015-CONVECS}} developed at INRIA Grenoble, the REC specification was automatically translated into various languages: AProVE (TRS), Clean, Haskell, LNT, LOTOS, Maude, mCRL2, OCaml, Opal, Rascal, Scala, SML, Stratego/XT, and Tom. Using the interpreters, compilers, and checkers available for these languages, it was shown \cite[Sect.~5]{Garavel-Marsso-17} that the REC specification terminates, that it is confluent, and that all the~203 tests pass successfully. Also, the most involved components (namely, the binary adders and multipliers) were validated separately using more than 30,000 test vectors.
\end{itemize}

\noindent
	The two remaining formal specifications of the MAA are novel contributions of the present paper:

\begin{itemize}
	\item \SPEC{LOTOS-17}\,: Between January and February 2017, Hubert Garavel and Lina Marsso performed a major revision of \SPEC{LOTOS-92} based upon the detailed knowledge of the MAA acquired during the development of \SPEC{REC-17}\,. Their goal was to produce an executable LOTOS specification as simple as possible, even if it departed from the original specification \SPEC{LOTOS-91} written by Harold B.~Munster.
	Many changes were brought: the two sorts \T{AcceptableMessage} and \T{SegmentedMessage} were removed, and the \T{Nat} sort was replaced almost everywhere by the \T{Block} sort; about seventy operations were removed, while a dozen new operations were added; the \T{Block} constructor evolved by taking four octets rather than thirty-two bytes; the constructors of sort \T{Message} were replaced by standard list constructors; the equations defining various operations (\T{FIX1}, \T{FIX2}, \T{BYT}, \T{PAT}, etc.) were shortened; each message is now processed in a single pass without first duplicating it to build a list of segments; the \T{Prelude} operation is executed only once per message, rather than once per segment; the detection of messages larger than 1,000,000 blocks is now written directly in C.
	These changes led to a 266-line LOTOS specification (see Annex~\ref{ANNEX-LOTOS}) with two companion C~files (157 lines in total) implementing the basic operations on blocks\footnote{The most recent version of these files is available from \url{ftp://ftp.inrialpes.fr/pub/vasy/demos/demo_12/LOTOS}.}. Interestingly, all these files taken together are smaller than the original specification \SPEC{LOTOS-91}\,, demonstrating that executability and conciseness are not necessarily antagonistic notions.

	\item \SPEC{LNT-17}\,: Between December 2016 and February 2017, Hubert Garavel and Lina Marsso entirely rewrote \SPEC{LNT-16} in order to obtain a simpler specification. First, the same changes as for \SPEC{LOTOS-17} were applied to the LNT specification.
	Also, the sorts \T{Pair}, \T{TwoPairs}, and \T{ThreePairs}, which had been introduced by Harold B.~Munster to describe functions returning two, four, and six blocks, have been eliminated; this was done by having LNT functions that return their computed results using ``\B{out}'' or ``\B{in out}'' parameters (i.e., call by result or call by value-result) rather than tuples of values;
	the principal functions (e.g., \T{MUL1}, \T{MUL2}, \T{MUL2A}, \T{Prelude}, \T{Coda}, \T{MAC}, etc.) have been simplified by taking advantage of the imperative style LNT, i.e., mutable variables and assignments;
	many auxiliary functions have been gathered and replaced by a few larger functions (e.g., \T{PreludeJ}, \T{PreludeK}, \T{PreludeHJ}, and \T{PreludeHK}) also  written in the imperative style.
	These changes resulted in a 268-line LNT specification with a 136-line companion C~file, which have nearly the same size as \SPEC{LOTOS-17}\,, although the LNT version is more readable and closer to the original MAA specification \cite{Davies-Clayden-88}, also expressed in the imperative style. Taken alone, the LNT code has approximately the same size as \SPEC{VDM-90}\,, the non-executable specification that was included as a formal annex in the MAA standard \cite{ISO-8731-2:1992}.

	As for \SPEC{REC-17}\,, the LNT specification was then enriched with a collection of ``\B{assert}'' statements implementing: (i) the test vectors listed in Tables~1 to~6 of \cite[Annex~A]{ISO-8731-2:1992} and \cite{Davies-Clayden-88}; (ii) the test vectors of \cite[Annex~E.3.3]{ISO-8730:1990}; (iii) supplementary test vectors intended to specifically check for certain aspects (byte permutations and message segmentation) that were not enough covered by the above tests; this was done by introducing a \T{makeMessage} function acting as a pseudo-random message generator.

	Consequently, the size of the LNT files grew up to 1334 lines in total (see Annex~\ref{ANNEX-LNT})\footnote{The most recent version of these files is available from \url{ftp://ftp.inrialpes.fr/pub/vasy/demos/demo_12}.}.
	Finally, the remaining test vectors of \cite[Annexes~E.3.4 and~E.4]{ISO-8730:1990}, which were too lengthy to be included in \SPEC{REC-17}\,, have been stored in text files and can be checked by running the C~code generated from the LNT specification. This makes of \SPEC{LNT-17} the most complete formal specification of the MAA as far as validation is concerned.

\end{itemize}

\section{Modelling issues}
\label{DISCUSSION}

	In this section, we investigate some salient issues faced when modelling the MAA using diverse formal methods. We believe that such issues are not specific to the MAA, but are likely to arise whenever non-trivial data structures and algorithms are to be described formally.

\subsection{Local variables in function definitions}

	Local variables are essential to store computed results that need to be used several times, thus avoiding identical calculations to be repeated. LNT allows to freely define and assign local variables in an imperative-programming style; the existence of a formal semantics is guaranteed by static semantic constraints \cite{Garavel-15-b} ensuring that each variable is duly assigned before used. For instance, the \T{MUL1} function\footnote{The same discussion is also valid for \T{MUL2}, \T{MUL2A}, and many other MAA functions.} is expressed in LNT as follows:

\begin{small}
\begin{verbatim}
   function MUL1 (X, Y : Block) : Block is
      var U, L, S, C : Block in
         U := HIGH_MUL (X, Y);
         L := LOW_MUL (X, Y);
         S := ADD (U, L);
         C := CAR (U, L);
         assert (C == x00000000) or (C == x00000001);
         return ADD (S, C)
      end var
   end function
\end{verbatim}
\end{small}

\noindent In VDM, which enjoys a ``\B{let}'' operator, the definition of \T{MUL1} is very similar to the LNT one \cite[page~11]{Parkin-ONeill-90} \cite[Sect.~2.2.5]{Parkin-ONeill-91}. The situation is quite different for term rewrite systems and abstract data types, which lack a ``\B{let}'' operator in their rewrite rules or equations. Interestingly, \SPEC{LOTOS-91} tries to emulate such a ``\B{let}'' operator by (ab)using the premises of conditional equations \cite[pages~37 and~78]{Munster-91-a}:

\begin{small}
\begin{verbatim}
   opns MUL1 : Block, Block -> Block
   forall X, Y, U, L, S, P: Block, C: Bit
      NatNum (X) * NatNum (Y) = NatNum (U ++ L),
      NatNum (U) + NatNum (L) = NatNum (S) + NatNum (C),
      NatNum (C + S) = NatNum (P)
      => MUL1 (X, Y) = P;
\end{verbatim}
\end{small}

\noindent These premises define and compute\footnote{These premises silently require the computation of inverse functions for \T{NumNat}, \T{+}, and \T{++} (bit string concatenation).} the variables (\T{U}, \T{L}), (\T{S}, \T{C}), and \T{P}, respectively. Unfortunately, most languages and tools for term rewriting forbid such free variables in premises, requiring that only the parameters of the function under definition (here, \T{X} and \T{Y} for the \T{MUL1} function) can occur in premises.

	Instead, \SPEC{LOTOS-17} and \SPEC{REC-17} adopt a more conventional style in which auxiliary operations are introduced, the parameters of which are used to store computed results that need to be used more than once:

\begin{small}
\begin{verbatim}
   opns MUL1    : Block, Block -> Block
        MUL1_UL : Block, Block -> Block
        MUL1_SC : Block, Block -> Block
   forall X, Y, U, L, S, C : Block
      MUL1 (X, Y)    = MUL1_UL (HIGH_MUL (X, Y), LOW_MUL (X, Y));
      MUL1_UL (U, L) = MUL1_SC (ADD (U, L), CAR (U, L));
      MUL1_SC (S, C) = ADD (S, C);
\end{verbatim}
\end{small}

\noindent In comparison, the imperative-programming style of LNT is clearly more concise, more readable, and closer to the original description of \T{MUL1}. Moreover, LNT permits successive assignments to the same variable, which proved to be useful in, e.g., the \T{MainLoop} and \T{MAC} functions.

\subsection{Functions returning multiple results}

Another point in which the various MAA specifications differ is the handling of functions that compute more than one result. There are several such functions in the MAA; let us consider the \T{Prelude} function, which takes two block parameters \T{J} and \T{K} and returns six block parameters \T{X}, \T{Y}, \T{V}, \T{W}, \T{S}, and \T{T}.

The simplest description of this function is achieved in \SPEC{LNT-17}, which exploits the fact that LNT functions, like in imperative programming languages, may return a result and/or have ``\B{out}'' parameters. In LNT, the \T{Prelude} function can be defined this way:
\begin{small}
\begin{verbatim}
   function Prelude (in J, K : Block, out X, Y, V, W, S, T : Block) is
      ...
   end function
\end{verbatim}
\end{small}
and invoked as follows:
\begin{small}
\begin{verbatim}
   Prelude (J, K, ?X0, ?Y0, ?V0, ?W, ?S, ?T)
\end{verbatim}
\end{small}

\noindent Although this approach is the simplest one, most formal methods do not support procedures or functions with ``\B{out}'' parameters\footnote{Besides LNT, the only other language we know to offer ``\B{out}'' parameters is the synchronous dataflow language Lustre.}. In such languages where functions return only a single result, there are two different options for describing functions with multiple results such as \T{Prelude}.

	The first option is return a unique result of some compound type (record, tuple, array, etc.). For instance, both \SPEC{VDM-90} and \SPEC{Z-91} describe \T{Prelude} as a function taking a pair of blocks and returning a result of a new type (called \I{Key-Constant} \cite[Sections~2.2.2 and 2.2.7]{Parkin-ONeill-90} or \I{DerivedSpace} \cite[pages 45--46]{Lai-91}) defined as a sextuple of blocks.
	\SPEC{LOTOS-91} and \SPEC{LOTOS-17} adopt a similar approach by defining \T{Prelude} to return a result of a new sort \T{ThreePairs}, which is a triple of \T{Pair} values, where sort \T{Pair} is itself defined as a pair of blocks.
	Other examples can be found in the binary adders and multipliers of \SPEC{REC-17}\,; for instance, the 8-bit adder returns a result of sort \T{OctetSum} that is a pair gathering a sum (of sort \T{Octet}) and a carry (of sort \T{Sum}).

	The drawbacks of this first option are numerous: (i) new types have to be introduced --- potentially one type per defined function in the worst case; (ii) each of these types introduces in turn a constructor and, often, equality and projection functions as well; (iii) the specification gets obscured by tupling/detupling operations, with the aggravating circumstance that detupling can be performed in different ways (pattern matching, destructuring ``\B{let}'', or projection functions), which makes it difficult to follow the flow of a particular variable embedded in a tuple of values; (iv) tupling complicates the efforts of compilers and garbage collector to allocate memory efficiently.

	The second option is to split a function returning $N>1$ results into $N$ separate functions. For instance, \SPEC{REC-17} has split \T{Prelude} into three operations: \T{preludeXY}, which computes the pair $(\T{X0}, \T{Y0})$, \T{preludeVW}, which computes the pair $(\T{V0}, \T{W})$, and \T{preludeST}, which computes the pair $(\T{S}, \T{T})$. This transformation applied to \T{Prelude} and to the main-loop functions enabled the sorts \T{TwoPairs} and \T{ThreePairs} introduced in \SPEC{LOTOS-91} to be entirely removed from \SPEC{REC-17}\,.

	The drawbacks of this second option are two-fold: (i) splitting a function with multiple results might be difficult if the calculations for these results are tightly intertwined; this was not the case with the six \T{Prelude} results, each of which does not depend on the five other ones\footnote{This was pointed out as a cryptographic weakness of the MAA in \cite[Sect.~6]{Rijmen-Preneel-DeWin-96}.}; (ii) splitting may require to duplicate identical calculations, and thus create inefficiencies that in turn may require the introduction of auxiliary functions to be avoided.

\section{Validation of MAA Specifications}
\label{VALIDATION}

	The two most recent specifications of the MAA have been validated as follows:

\begin{itemize}
	\item \SPEC{LOTOS-17}\,: The specification was validated by the \CAESARADT\ compiler, which implements all the syntactic and semantic checks stated in the definition of LOTOS \cite{ISO-8807}. The C~code generated from the LOTOS specification passed the test vectors specified in \cite[Annexes~E.3.4 and~E.4]{ISO-8730:1990}.

	\item \SPEC{LNT-17}\,: The specification was validated by the LNT2LOTOS translator, which implements the syntactic checks and (part of) the semantic checks stated in the definition of LNT \cite{Champelovier-Clerc-Garavel-et-al-10-v6.7} and generates LOTOS code, which is then validated by the \CAESARADT\ compiler, therefore performing the remaining semantics checks of LNT. The C code generated by the \CAESARADT\ compiler passed the test vectors specified in \cite[Annex~A]{ISO-8731-2:1992}, in \cite[Annexes~E.3]{ISO-8730:1990}, in \cite[Annexes~E.3.4 and~E.4]{ISO-8730:1990}, and the supplementary test vectors based on the \T{MakeMessage} function.
\end{itemize}

	Due to these checks, various mistakes were discovered in prior (informal and formal) specifications of the MAA:
	(i) Annex~\ref{ERRATA-8730} corrects the test vectors given in \cite[Annex~E]{ISO-8730:1990};
	(ii) Annex~\ref{ERRATA-8731} corrects the test vectors given for function PAT in \cite[Annex~A]{ISO-8731-2:1992} and \cite{Davies-Clayden-88};
	(iii) an error was found in the main~C program, which computed an incorrect MAC value, as the list of blocks storing the message was built in reverse order;
	(iv) another error was found in the external implementation in~C of the function \T{HIGH_MUL}, which computes the highest 32 bits of the 64-bit product of two blocks and is imported by the LOTOS and LNT specifications --- this illustrates the risks arising when formal and non-formal codes are mixed.

\section{Conclusion}
\label{CONCLUSION}

	Twenty-five years after, we revisited the Message Authenticator Algorithm (MAA), which used to be a pioneering case study for cryptography in the 80s and for formal methods in the early 90s.
	The three MAA specifications \SPEC{VDM-90}\,, \SPEC{Z-91}\,, and \SPEC{LOTOS-91} developed at NPL in 1990--1991 were clearly leading-edge, as can be seen from the adoption of the VDM specification as part of the ISO international standard 8731-2 in 1992. However, they also faced limitations: these were mostly pen-and-pencil formal methods that lacked automated validation tools and that required implementations to be developed manually, thus raising the difficult question of the compatibility between the formal specification and the handwritten implementation code.

	A different path has been followed at INRIA Grenoble since the early 90s, with an emphasis on executable formal methods, from which implementations can be generated automatically.
Five specifications have been successively developed: \SPEC{LOTOS-92}\,, \SPEC{LNT-16}\,, \SPEC{REC-17}\,, \SPEC{LOTOS-17}\,, and \SPEC{LNT-17}\,. Retrospectively, heading towards executable formal methods proved to be a successful bet:

\begin{itemize}
	\item It turns out that executable specifications are not necessarily longer than non-executable ones: \SPEC{LNT-17} and \SPEC{LOTOS-17} (345 and 423 lines, respectively, including the external C~code fragments) are half way between the non-executable specifications \SPEC{VDM-90} (275 lines) and \SPEC{Z-91} (608 lines). Also, \SPEC{LNT-17} is only 60\% larger than the direct implementation in C given in \cite{Davies-Clayden-88}.

	\item One might argue that the LOTOS and LNT specifications are not entirely formal, as they import a few C~types and functions to implement blocks and arithmetic operations on blocks. We see this as a strength, rather than a weakness, of our approach. Moreover, nothing prevents such external types and functions to be instead defined in LOTOS or in LNT, as this was the case with the \SPEC{REC-17} specification, which was then automatically translated to self-contained, fully-formal LOTOS and LNT specifications that were successfully compiled and executed.

	\item The insight gained by comparing the eight formal specifications of the MAA confirms that LNT is a formal method of choice for modelling complex algorithms and data structures. Compared to other formalisms, LNT offers an imperative specification style (based on mutable variables and assignments) that proved to be simpler to write, easier to read, more concise, and closer to the MAA description in natural language \cite{Davies-Clayden-88}, from which specifications based on term rewrite systems and abstract data types significantly depart due to picky technical restrictions in these latter formalisms.
	LNT also favors a more disciplined specification style that, we believe, is of higher quality because of the numerous static-analysis checks (e.g., unused variables, useless assignments, etc.) performed by the LNT2LOTOS translator; such strict controls are, to the best of our knowledge, absent from most other specification languages.

	\item The application of executable formal methods to the MAA case study was fruitful in several respects: (i) it detected errors in the reference test vectors given in ISO standards 8730 and 8731-2; (ii) the LOTOS specification of the MAA, due to its size and complexity, was helpful in improving early versions of the \CAESARADT\ compiler; (iii)~similarly, the LNT specification of the MAA revealed in the LNT2LOTOS translator a few defects and performance issues, which have been dealt with in 2016 and 2017.

	\item Moreover, executable formal methods benefit from significant progress in their compiling techniques. In 1990, a handwritten implementation of the MAA in Miranda took 60~seconds to process an 84-block message and 480 seconds to process a 588-block message \cite[page~37]{Parkin-ONeill-90}. Today, the implementations automatically generated from the LNT and LOTOS specifications of the MAA take~0.65 and 0.37 second, respectively, to process a one-million-block message\footnote{The C code generated from LNT and LOTOS by the CADP translators was compiled using ``\T{gcc -O3}'' and ran on a Dell Latitude E6530 laptop.}. As it appears, ``formal'' and ``executable'' are no longer mutually exclusive qualities.

\end{itemize}

\subsection*{Acknowledgements}

	We are grateful to Philippe Turlier who, in 1992, helped turning the non-executable LOTOS specification of Harold B. Munster into an executable one, to Wendelin Serwe, who, in 2016, produced the first LNT specification of the MAA, and to Fr\'{e}d\'{e}ric Lang, who, in 2016--2017, improved the LNT2LOTOS translator to address the issues pointed out.
	Acknowledgements are also due to Keith Lockstone for his advice and his web site\footnote{\url{http://www.cix.co.uk/~klockstone}} giving useful information about the MAA, and to Sharon Wilson, librarian of the National Physical Laboratory, who provided us with valuable early NPL reports that cannot be fetched from the web.

\appendix

\section{Errata Concerning Annex E of the ISO-8730:1990 Standard}
\label{ERRATA-8730}

	After reading and checking carefully the test vectors given in \cite[Annex~E]{ISO-8730:1990}, we discovered a number of errors\footnote{We used the French version of that standard, which we acquired from AFNOR, but have no reason to believe that the same errors are absent from other translations of this standard.}. Here is the list of errors found and their corrections:

\begin{itemize}
	\item In Annex~E.2, some characters of the text message differ from the corresponding ASCII code given (in hexadecimal) below in Annex E.3.2. Precisely, the string \verb+"BE CAREFUL"+ should read \verb+"BE\n\n\ \ \ Careful"+, where \verb+"\n"+ and \verb+"\ "+ respectively denote line-feed and white space. The corresponding hexadecimal values are indeed 42 45 0A 0A 20 20 20 43 61 72 65 66 75 6C.

	\item Annex~E.3.2 and Annex~E.3.4 state that this text message has 86 blocks. Actually, it has \underline{84} blocks only. This is confirmed by the table of hexadecimal values in Annex~E.3.2 (42~lines $\times$ 2 blocks per line give 84 blocks) and by the iterations listed in Annex~E.3.4, in which the number of message blocks (i.e., variable \T{N}) ranges between 1 and 84.

	\item Annex~E.4 states that the long message is obtained by repeating six times the message of 86~blocks, leading to a message length of 516~blocks. Actually, it is obtained by repeating \underline{seven} times the message of \underline{84}~blocks, leading to a message length of \underline{588}~blocks. This can be seen from the iterations listed in Annex~E.4 where variable \T{N} ranges between 1 and 588, and by the fact that $588 = 7 \times 84$. Moreover, computing the MAA result on the 588-block long message with the same key J~=~\T{E6 A1 2F 07} and K~=~\T{9D 15 C4 37} as in Annex~E.3.3 indeed gives the expected MAC value \T{C6 E3 D0 00}.
\end{itemize}

\section{Errata Concerning Annex A of the ISO-8731-2:1992 Standard}
\label{ERRATA-8731}

	After checking carefully all the test vectors contained in the original NPL report defining the MAA \cite{Davies-Clayden-88} and in the 1992 version of the MAA standard \cite{ISO-8731-2:1992}, we believe that there are mistakes\footnote{Again, we used the French version of this standard, but we believe that this plays no role, as the same mistakes were already present in the 1988 NPL report.} in the test vectors given for function \T{PAT}.

	More precisely, the three last lines of Table~3 \cite[page 15]{Davies-Clayden-88} --- identically reproduced in Table~A.3 of \cite[Sect.~A.4]{ISO-8731-2:1992} --- are written as follows:

\begin{small}
\begin{verbatim}
	{X0,Y0}     0103 0703 1D3B 7760     PAT{X0,Y0} EE
	{V0,W}      0103 050B 1706 5DBB     PAT{V0,W}  BB
	{S,T}       0103 0705 8039 7302     PAT{S,T}   E6
\end{verbatim}
\end{small}

	Actually, the inputs of function \T{PAT} should not be \verb+{X0,Y0}+, \verb+{V0,W}+, \verb+{S,T}+ but rather \verb+{H4,H5}+, \verb+{H6,H7}+, \verb+{H8,H9}+, the values of \T{H4}, ..., \T{H9} being those listed above in Table~3. Notice that the confusion was probably caused by the following algebraic identities:

\begin{small}
\begin{verbatim}
	{X0,Y0} = BYT (H4, H5)
	{V0,W}  = BYT (H6, H7)
	{S,T}   = BYT (H8, H9)
\end{verbatim}
\end{small}

	If one gives \verb+{X0,Y0}+, \verb+{V0,W}+, \verb+{S,T}+ as inputs to \T{PAT}, then the three results of \T{PAT} are equal to \T{00} and thus cannot be equal to \T{EE}, \T{BB}, \T{E6}, respectively.

	But if one gives \verb+{H4,H5}+, \verb+{H6,H7}+, \verb+{H8,H9}+ as inputs to \T{PAT}, then the results of \T{PAT} are the expected values \T{EE}, \T{BB}, \T{E6}.

	Thus, we believe that the three last lines of Table 3 should be modified as follows:

\begin{small}
\begin{verbatim}
	{H4,H5}     0000 0003 0000 0060     PAT{H4,H5}  EE
	{H6,H7}     0003 0000 0006 0000     PAT{H6,H7}  BB
	{H8,H9}     0000 0005 8000 0002     PAT{H8,H9}  E6
\end{verbatim}
\end{small}

\clearpage

\section{Formal Specification of the MAA in LOTOS}
\label{ANNEX-LOTOS}

\lstset{
  language=LOTOS,
  basicstyle=\ttfamily\small,
  columns=fullflexible, 
  commentstyle=\rmfamily\itshape,
  keywordstyle=\rmfamily\bfseries,
  xleftmargin=\parindent
}

This annex presents the specification \SPEC{LOTOS-17} of the MAA in LOTOS. This specification uses several predefined libraries of LOTOS, namely: the libraries for Booleans and natural numbers, which we do not reproduce here, and the libraries for bits, octets, and octet values, of which we only display excerpts needed for understanding the MAA specification.

\subsection{The BIT library}

This predefined LOTOS library defines the \T{Bit} type with its related operations. Only a simplified version of this library is presented here.

\begin{lstlisting}
type Bit is Boolean, NaturalNumber
   sorts
      Bit

   opns
      0 (*! constructor *), 1 (*! constructor *) : -> Bit
      not : Bit -> Bit
      _and_, _or_, _xor_ : Bit, Bit -> Bit
      _eq_, _ne_ : Bit, Bit -> Bool

   eqns

   forall x,y : Bit
   ofsort Bit

      not (0) = 1;
      not (1) = 0;

      x and 0 = 0;
      x and 1 = x;

      x or 0 = x;
      x or 1 = 1;

      x xor 0 = x;
      x xor 1 = not (x);

   ofsort Bool

      x eq x = true;
      0 eq 1 = false;
      1 eq 0 = false;

      x ne y = x eq y;
endtype

\end{lstlisting}

\subsection{The OCTET library}

This predefined LOTOS library defines the \T{Octet} type (i.e., an 8-bit word) with its related operations. Only an excerpt of this library is presented here.

\begin{lstlisting}
type Octet is Bit, Boolean
   sorts
      Octet

   opns
      Octet (*! constructor *) : Bit, Bit, Bit, Bit, Bit, Bit, Bit, Bit -> Octet
      Bit1, Bit2, Bit3, Bit4, Bit5, Bit6, Bit7, Bit8 : Octet -> Bit
      not : Octet -> Octet
      _and_, _or_, _xor_ : Octet, Octet -> Octet
      _eq_, _ne_ : Octet, Octet -> Bool

   eqns

   ofsort Bit
   forall b1, b2, b3, b4, b5, b6, b7, b8 : Bit

      Bit1 (Octet (b1, b2, b3, b4, b5, b6, b7, b8)) = b1;
      Bit2 (Octet (b1, b2, b3, b4, b5, b6, b7, b8)) = b2;
      Bit3 (Octet (b1, b2, b3, b4, b5, b6, b7, b8)) = b3;
      Bit4 (Octet (b1, b2, b3, b4, b5, b6, b7, b8)) = b4;
      Bit5 (Octet (b1, b2, b3, b4, b5, b6, b7, b8)) = b5;
      Bit6 (Octet (b1, b2, b3, b4, b5, b6, b7, b8)) = b6
      Bit7 (Octet (b1, b2, b3, b4, b5, b6, b7, b8)) = b7;
      Bit8 (Octet (b1, b2, b3, b4, b5, b6, b7, b8)) = b8;

   ofsort Octet
   forall x, y : Octet

      not (x) = Octet (not (Bit1 (x)), not (Bit2 (x)),
                       not (Bit3 (x)), not (Bit4 (x)), 
                       not (Bit5 (x)), not (Bit6 (x)), 
                       not (Bit7 (x)), not (Bit8 (x)));

      x and y = Octet (Bit1 (x) and Bit1 (y), Bit2 (x) and Bit2 (y),
                       Bit3 (x) and Bit3 (y), Bit4 (x) and Bit4 (y), 
                       Bit5 (x) and Bit5 (y), Bit6 (x) and Bit6 (y), 
                       Bit7 (x) and Bit7 (y), Bit8 (x) and Bit8 (y));

      x or y =  Octet (Bit1 (x) or Bit1 (y), Bit2 (x) or Bit2 (y),
                       Bit3 (x) or Bit3 (y), Bit4 (x) or Bit4 (y),
                       Bit5 (x) or Bit5 (y), Bit6 (x) or Bit6 (y), 
                       Bit7 (x) or Bit7 (y), Bit8 (x) or Bit8 (y));

      x xor y = Octet (Bit1 (x) xor Bit1 (y), Bit2 (x) xor Bit2 (y),
                       Bit3 (x) xor Bit3 (y), Bit4 (x) xor Bit4 (y), 
                       Bit5 (x) xor Bit5 (y), Bit6 (x) xor Bit6 (y), 
                       Bit7 (x) xor Bit7 (y), Bit8 (x) xor Bit8 (y));

   ofsort Bool
   forall x, y : Octet

      x eq y = (Bit1 (x) eq Bit1 (y)) and (Bit2 (x) eq Bit2 (y)) and
               (Bit3 (x) eq Bit3 (y)) and (Bit4 (x) eq Bit4 (y)) and
               (Bit5 (x) eq Bit5 (y)) and (Bit6 (x) eq Bit6 (y)) and
               (Bit7 (x) eq Bit7 (y)) and (Bit8 (x) eq Bit8 (y));

      x ne y = not (x eq y);
endtype
\end{lstlisting}

\subsection{The OCTETVALUES library}

This predefined LOTOS library defines 256 constant functions \T{x00}, ..., \T{xFF} that provide shorthand notations for octet values. Only an excerpt of this library is presented here.

\begin{lstlisting}
type OctetValues is Bit, Octet
   opns
      x00, x01, ... xFE, xFF : -> Octet

   eqns

   ofsort Octet

      x00 = Octet (0, 0, 0, 0, 0, 0, 0, 0);
      x01 = Octet (0, 0, 0, 0, 0, 0, 0, 1);
      ...
      xFE = Octet (1, 1, 1, 1, 1, 1, 1, 0);
      xFF = Octet (1, 1, 1, 1, 1, 1, 1, 1);
endtype
\end{lstlisting}

\subsection{The MAA specification}

\begin{lstlisting}
specification MAA: noexit

library
   X_NATURAL, BIT, BOOLEAN, OCTET, OCTETVALUES
endlib

(* ----------------------------------------------------- *)

type Block is Boolean, Bit, Octet
   sorts
      Block (*! implementedby BLOCK printedby PRINT_BLOCK *)

   opns
      Block (*! implementedby BUILD_BLOCK constructor *) :
         Octet, Octet, Octet, Octet -> Block

      _eq_ (*! implementedby EQUAL_BLOCK *) : Block, Block -> Bool

      AND : Block, Block -> Block
      OR  : Block, Block -> Block
      XOR : Block, Block -> Block
      CYC : Block -> Block

      ADD      (*! implementedby ADD external *)      : Block, Block -> Block
      CAR      (*! implementedby CAR external *)      : Block, Block -> Block
      HIGH_MUL (*! implementedby HIGH_MUL external *) : Block, Block -> Block
      LOW_MUL  (*! implementedby LOW_MUL external *)  : Block, Block -> Block

   eqns

   ofsort Bool
   forall X, Y : Block

      X eq X = true;
      X eq Y = false; (* assuming priority between equations *)

   ofsort Block
   forall O1, O2, O3, O4, P1, P2, P3, P4 : Octet

      AND (Block (O1, O2, O3, O4), Block (P1, P2, P3, P4)) =
         Block (O1 and P1, O2 and P2, O3 and P3, O4 and P4);

      OR (Block (O1, O2, O3, O4), Block (P1, P2, P3, P4)) =
         Block (O1 or P1, O2 or P2, O3 or P3, O4 or P4);

      XOR (Block (O1, O2, O3, O4), Block (P1, P2, P3, P4)) =
         Block (O1 xor P1, O2 xor P2, O3 xor P3, O4 xor P4);

   ofsort Block
   forall B1,  B2,  B3,  B4,  B5,  B6,  B7,  B8,
          B9,  B10, B11, B12, B13, B14, B15, B16,
          B17, B18, B19, B20, B21, B22, B23, B24,
          B25, B26, B27, B28, B29, B30, B31, B32 : Bit

      CYC (Block (Octet (B1,  B2,  B3,  B4,  B5,  B6,  B7,  B8),
                  Octet (B9,  B10, B11, B12, B13, B14, B15, B16),
                  Octet (B17, B18, B19, B20, B21, B22, B23, B24),
                  Octet (B25, B26, B27, B28, B29, B30, B31, B32))) =
         Block (Octet (B2,  B3,  B4,  B5,  B6,  B7,  B8,  B9),
                Octet (B10, B11, B12, B13, B14, B15, B16, B17),
                Octet (B18, B19, B20, B21, B22, B23, B24, B25),
                Octet (B26, B27, B28, B29, B30, B31, B32, B1));
endtype

(* ----------------------------------------------------- *)

type Pairs is Block
   sorts
      Pair,
      TwoPairs,
      ThreePairs

   opns
      Pair (*! constructor *) : Block, Block -> Pair
      TwoPairs (*! constructor *) : Pair, Pair -> TwoPairs
      ThreePairs (*! constructor *) : Pair, Pair, Pair -> ThreePairs
endtype

(* ----------------------------------------------------- *)

type Message is Block, Natural
   sorts
      Message (*! implementedby MESSAGE *)

   opns
      nil (*! implementedby NIL_MESSAGE constructor *) : -> Message
      _++_ (*! implementedby PLUS_MESSAGE constructor *) :
         Block, Message  -> Message

      Reverse (*! implementedby REVERSE external *) : Message  -> Message
      (* Reverse is not invoked in "maa.lotos" but in "main.c" *)
      (* for efficiency reason, Reverse is implemented directly in C *)
endtype

(* ----------------------------------------------------- *)

type Functions is Block, OctetValues, Pairs
   opns
      A_CONSTANT : -> Block
      B_CONSTANT : -> Block
      C_CONSTANT : -> Block
      D_CONSTANT : -> Block

      FIX1 : Block -> Block
      FIX2 : Block -> Block

      MUL1     : Block, Block -> Block
      MUL1_UL  : Block, Block -> Block
      MUL1_SC  : Block, Block -> Block

      MUL2     : Block, Block -> Block
      MUL2_UL  : Block, Block -> Block
      MUL2_DEL : Block, Block, Block -> Block
      MUL2_FL  : Block, Block -> Block
      MUL2_SC  : Block, Block -> Block

      MUL2A    : Block, Block -> Block
      MUL2A_UL : Block, Block -> Block
      MUL2A_DL : Block, Block -> Block
      MUL2A_SC : Block, Block -> Block

      NeedAdjust : Octet -> Bool
      AdjustCode : Octet -> Bit
      Adjust     : Octet, Octet -> Octet

      PAT        : Block, Block -> Octet
      BYT        : Block, Block -> Pair
      AuxBYT     : Block, Block, Octet -> Pair

   eqns

   ofsort Block

      A_CONSTANT = Block (x02, x04, x08, x01);
      B_CONSTANT = Block (x00, x80, x40, x21);
      C_CONSTANT = Block (xBF, xEF, x7F, xDF);
      D_CONSTANT = Block (x7D, xFE, xFB, xFF);

   ofsort Block
   forall X : Block

      FIX1 (X) = AND (OR (X, A_CONSTANT), C_CONSTANT);
      FIX2 (X) = AND (OR (X, B_CONSTANT), D_CONSTANT);

   ofsort Block
   forall X, Y, U, L, S, C : Block

      MUL1 (X, Y)    = MUL1_UL (HIGH_MUL (X, Y), LOW_MUL (X, Y));
      MUL1_UL (U, L) = MUL1_SC (ADD (U, L), CAR (U, L));
      MUL1_SC (S, C) = ADD (S, C);

   ofsort Block
   forall X, Y, U, L, D, F, E, S, C : Block

      MUL2 (X, Y)        = MUL2_UL (HIGH_MUL (X, Y), LOW_MUL (X, Y));
      MUL2_UL (U, L)     = MUL2_DEL (ADD (U, U), CAR (U, U), L);
      MUL2_DEL (D, E, L) = MUL2_FL (ADD (D, ADD (E, E)), L);
      MUL2_FL (F, L)     = MUL2_SC (ADD (F, L), CAR (F, L));
      MUL2_SC (S, C)     = ADD (S, ADD (C, C));

   ofsort Block
   forall X, Y, U, L, D, S, C : Block

      MUL2A (X, Y)    = MUL2A_UL (HIGH_MUL (X, Y), LOW_MUL (X, Y));
      MUL2A_UL (U, L) = MUL2A_DL (ADD (U, U), L);
      MUL2A_DL (D, L) = MUL2A_SC (ADD (D, L), CAR (D, L));
      MUL2A_SC (S, C) = ADD (S, ADD (C, C));

   ofsort Bool
   forall O: Octet

      NeedAdjust (O) = (O eq x00) or (O eq xFF);

   ofsort Bit
   forall O: Octet

      NeedAdjust (O)       => AdjustCode (O) = 1;
      not (NeedAdjust (O)) => AdjustCode (O) = 0;

   ofsort Octet
   forall O, P: Octet

      NeedAdjust (O)       => Adjust (O, P) = O xor P;
      not (NeedAdjust (O)) => Adjust (O, P) = O;

   ofsort Octet
   forall X, Y: Block, O1, O2, O3, O4, O5, O6, O7, O8: Octet

      PAT (Block (O1, O2, O3, O4), Block (O5, O6, O7, O8)) =
         Octet (AdjustCode (O1), AdjustCode (O2),
                AdjustCode (O3), AdjustCode (O4),
                AdjustCode (O5), AdjustCode (O6),
                AdjustCode (O7), AdjustCode (O8));

   ofsort Pair
   forall B1, B2, B3, B4, B5, B6, B7, B8 : Bit,
          O1, O2, O3, O4, O5, O6, O7, O8: Octet,
          J, K : Block

      BYT (J, K) = AuxBYT (J, K, PAT (J, K));

      AuxBYT (Block (O1, O2, O3, O4), Block (O5, O6, O7, O8),
              Octet (B1, B2, B3, B4, B5, B6, B7, B8)) =
         Pair (Block (Adjust (O1, Octet (0, 0, 0, 0, 0, 0, 0, B1)),
                      Adjust (O2, Octet (0, 0, 0, 0, 0, 0, B1, B2)),
                      Adjust (O3, Octet (0, 0, 0, 0, 0, B1, B2, B3)),
                      Adjust (O4, Octet (0, 0, 0, 0, B1, B2, B3, B4))),
               Block (Adjust (O5, Octet (0, 0, 0, B1, B2, B3, B4, B5)),
                      Adjust (O6, Octet (0, 0, B1, B2, B3, B4, B5, B6)),
                      Adjust (O7, Octet (0, B1, B2, B3, B4, B5, B6, B7)),
                      Adjust (O8, Octet (B1, B2, B3, B4, B5, B6, B7, B8))));
endtype

(* ----------------------------------------------------- *)

type Prelude is Functions
   opns
      Q                            : Octet -> Block
      SQUARE                       : Block -> Block
      J1_2, J1_4, J1_6, J1_8       : Block -> Block
      J2_2, J2_4, J2_6, J2_8       : Block -> Block
      K1_2, K1_4, K1_5, K1_7, K1_9 : Block -> Block
      K2_2, K2_4, K2_5, K2_7, K2_9 : Block -> Block
      H4, H6, H8, H0, H7, H9       : Block -> Block
      H5                           : Block, Octet -> Block
      Prelude    : Block, Block -> ThreePairs
      AuxPrelude : Pair, Octet -> ThreePairs

   eqns

   ofsort Block
   forall P: Octet

      Q (P) = SQUARE (ADD (Block (x00, x00, x00, P),
                           Block (x00, x00, x00, x01)));

   ofsort Block
   forall B: Block

      SQUARE (B) = LOW_MUL (B, B);

   ofsort Block
   forall J: Block

      J1_2 (J) = MUL1 (J, J);
      J1_4 (J) = MUL1 (J1_2 (J), J1_2 (J));
      J1_6 (J) = MUL1 (J1_2 (J), J1_4 (J));
      J1_8 (J) = MUL1 (J1_2 (J), J1_6 (J));
      J2_2 (J) = MUL2 (J, J);
      J2_4 (J) = MUL2 (J2_2 (J), J2_2 (J));
      J2_6 (J) = MUL2 (J2_2 (J), J2_4 (J));
      J2_8 (J) = MUL2 (J2_2 (J), J2_6 (J));

   ofsort Block
   forall K: Block

      K1_2 (K) = MUL1 (K, K);
      K1_4 (K) = MUL1 (K1_2 (K), K1_2 (K));
      K1_5 (K) = MUL1 (K, K1_4 (K));
      K1_7 (K) = MUL1 (K1_2 (K), K1_5 (K));
      K1_9 (K) = MUL1 (K1_2 (K), K1_7 (K));
      K2_2 (K) = MUL2 (K, K);
      K2_4 (K) = MUL2 (K2_2 (K), K2_2 (K));
      K2_5 (K) = MUL2 (K, K2_4 (K));
      K2_7 (K) = MUL2 (K2_2 (K), K2_5 (K));
      K2_9 (K) = MUL2 (K2_2 (K), K2_7 (K));

   ofsort Block
   forall J, K: Block, P: Octet

      H4 (J)    = XOR (J1_4 (J), J2_4 (J));
      H6 (J)    = XOR (J1_6 (J), J2_6 (J));
      H8 (J)    = XOR (J1_8 (J), J2_8 (J));

      H0 (K)    = XOR (K1_5 (K), K2_5 (K));
      H5 (K, P) = MUL2 (H0 (K), Q (P));
      H7 (K)    = XOR (K1_7 (K), K2_7 (K));
      H9 (K)    = XOR (K1_9 (K), K2_9 (K));

   ofsort ThreePairs
   forall J, K: Block, P: Octet

      Prelude (J, K) = AuxPrelude (BYT (J, K), PAT (J, K));

   ofsort ThreePairs
   forall J, K: Block, P: Octet

      AuxPrelude (Pair (J, K), P) =
         ThreePairs (BYT (H4 (J), H5 (K, P)),
                     BYT (H6 (J), H7 (K)),
                     BYT (H8 (J), H9 (K)));
endtype

(* ----------------------------------------------------- *)

type MAA is Prelude, Message
   opns
      ComputeXY : Pair, Block, Block -> Pair
      MainLoop  : TwoPairs, Block -> TwoPairs
      ComputeZ  : TwoPairs -> Block
      Coda      : TwoPairs, Pair -> Block
      255       (*! implementedby N255 external *) : -> Nat

      MAC      (*! implementedby MAC *) : Block, Block, Message -> Block
      MAAstart : ThreePairs, Message -> Block
      MAA      : Message, Nat, TwoPairs, Pair, Pair, Pair -> Block
      MAAjump  : Message, Block, Pair, Pair, Pair -> Block

   eqns

   ofsort Pair
   forall X, Y, M, E: Block

      ComputeXY (Pair (X, Y), M, E) =
         Pair (MUL1  (XOR (X, M), FIX1 (ADD (XOR (Y, M), E))),
               MUL2A (XOR (Y, M), FIX2 (ADD (XOR (X, M), E))));

   ofsort TwoPairs
   forall XY: Pair, B, V, W: Block

      MainLoop (TwoPairs (XY, Pair (V, W)), B) =
         TwoPairs (ComputeXY (XY, B, XOR (CYC (V), W)), Pair (CYC (V), W));

   ofsort Block
   forall X, Y: Block, VW: Pair

      ComputeZ (TwoPairs (Pair (X, Y), VW)) = XOR (X, Y);

   ofsort Block
   forall XYVW: TwoPairs, S, T: Block

      Coda (XYVW, Pair (S, T)) = ComputeZ (MainLoop (MainLoop (XYVW, S), T));

   ofsort Block
   forall J, K : Block, M: Message

      MAC (J, K, M) = MAAstart (Prelude (J, K), M);

   ofsort Block
   forall X0Y0, V0W, ST : Pair, M: Message

      MAAstart (ThreePairs (X0Y0, V0W, ST), M) =
         MAA (M, 255, TwoPairs (X0Y0, V0W), X0Y0, V0W, ST);

   ofsort Block
   forall X0Y0, V0W, ST : Pair, XYVW: TwoPairs, B : Block, N: Nat, M: Message

      MAA (nil, N, XYVW, X0Y0, V0W, ST) = Coda (XYVW, ST);
      MAA (B ++ M, 0, XYVW, X0Y0, V0W, ST) =
         MAAjump (M, Coda (MainLoop (XYVW, B), ST), X0Y0, V0W, ST);
      MAA (B ++ M, succ (N), XYVW, X0Y0, V0W, ST) =
         MAA (M, N, MainLoop (XYVW, B), X0Y0, V0W, ST);

   ofsort Block
   forall X0Y0, V0W, ST : Pair, B, Z : Block, M: Message

      MAAjump (nil, Z, X0Y0, V0W, ST) = Z;
      MAAjump (B ++ M, Z, X0Y0, V0W, ST) =
         MAA (B ++ M, 255, MainLoop (TwoPairs (X0Y0, V0W), Z), X0Y0, V0W, ST);
endtype

(* ----------------------------------------------------- *)

behaviour

    stop

endspec

\end{lstlisting}

\clearpage

\section{Formal Specification of the MAA in LNT}
\label{ANNEX-LNT}

\lstset{
  language=LNT,
  basicstyle=\ttfamily\small,
  columns=fullflexible, 
  commentstyle=\rmfamily\itshape,
  keywordstyle=\rmfamily\bfseries,
  xleftmargin=\parindent
}

This annex presents the specification \SPEC{LNT-17} of the MAA in LNT. This specification uses several predefined libraries of LNT, namely: the libraries for Booleans and natural numbers, which we do not reproduce here, and the libraries for bits, octets, and octet values, of which we only display excerpts needed for understanding the MAA specification. It also defines two new libraries for blocks and block values, which we display hereafter.

\subsection{The BIT library}

This predefined LNT library defines the \T{Bit} type with its related operations. Only an excerpt of this library is presented here.

\begin{lstlisting}
module BIT is

type BIT is
   0, 1
   with "eq", "ne", "lt", "le", "ge", "gt"
end type

function not (B : Bit) : Bit is
   case B in
     0 -> return 1
   | 1 -> return 0
   end case
end function

function _and_ (B1, B2 : Bit) : Bit is
   case B1 in
     0 -> return 0
   | 1 -> return B2
   end case
end function

function _or_ (B1, B2 : Bit) : Bit is
   case B1 in
     0 -> return B2
   | 1 -> return 1
   end case
end function

function _xor_ (B1, B2 : Bit) : Bit is
   case B1 in
     0 -> return B2
   | 1 -> return not (B2)
   end case
end function

end module
\end{lstlisting}

\subsection{The OCTET library}

This predefined LNT library defines the \T{Octet} type (i.e., an 8-bit word) with its related operations. Only an excerpt of this library is presented here.

\begin{lstlisting}
module OCTET (BIT) is

type Octet is
   Octet (B1, B2, B3, B4, B5, B6, B7, B8 : Bit)
   with "eq", "ne", "get"   
end type

function not (O : Octet) : Octet is
   return Octet (not (O.B1), not (O.B2),
                 not (O.B3), not (O.B4),
                 not (O.B5), not (O.B6),
                 not (O.B7), not (O.B8))
end function

function _and_ (O1, O2 : Octet) : Octet is
   return Octet (O1.B1 and O2.B1, O1.B2 and O2.B2,
                 O1.B3 and O2.B3, O1.B4 and O2.B4,
                 O1.B5 and O2.B5, O1.B6 and O2.B6,
                 O1.B7 and O2.B7, O1.B8 and O2.B8)
end function

function _or_ (O1, O2 : Octet) : Octet is
   return Octet (O1.B1 or O2.B1, O1.B2 or O2.B2,
                 O1.B3 or O2.B3, O1.B4 or O2.B4,
                 O1.B5 or O2.B5, O1.B6 or O2.B6,
                 O1.B7 or O2.B7, O1.B8 or O2.B8)
end function

function _xor_ (O1, O2 : Octet) : Octet is
   return Octet (O1.B1 xor O2.B1, O1.B2 xor O2.B2,
                 O1.B3 xor O2.B3, O1.B4 xor O2.B4,
                 O1.B5 xor O2.B5, O1.B6 xor O2.B6,
                 O1.B7 xor O2.B7, O1.B8 xor O2.B8)
end function

end module

\end{lstlisting}

\subsection{The OCTETVALUES library}

This predefined LNT library defines 256 constant functions \T{x00}, ..., \T{xFF} that provide shorthand notations for octet values. Only an excerpt of this library is presented here.

\begin{lstlisting}
module OCTETVALUES (BIT, OCTET) is

function x00 : Octet is
   return Octet (0, 0, 0, 0, 0, 0, 0, 0)
end function

function x01 : Octet is
   return Octet (0, 0, 0, 0, 0, 0, 0, 1)
end function
...
function xFE : Octet is
   return Octet (1, 1, 1, 1, 1, 1, 1, 0)
end function

function xFF : Octet is
   return Octet (1, 1, 1, 1, 1, 1, 1, 1)
end function

end module

\end{lstlisting}

\subsection{The BLOCK library}

This library defines the \T{Block} type (i.e., a 32-bit word) with its logical and arithmetical operations, the latter being implemented externally as a set of functions written in the C language.

\begin{lstlisting}
module BLOCK (BIT, OCTET) is

type Block is
   Block (O1, O2, O3, O4 : Octet)
   with "get", "=="
end type

function _==_ (O1, O2 : Octet) : Bool is
   return eq (O1, O2)
end function

function _AND_ (X, Y : Block) : Block is
   return Block (X.O1 and Y.O1, X.O2 and Y.O2, X.O3 and Y.O3, X.O4 and Y.O4)
end function

function _OR_ (X, Y : Block) : Block is
   return Block (X.O1 or Y.O1, X.O2 or Y.O2, X.O3 or Y.O3, X.O4 or Y.O4)
end function

function _XOR_ (X, Y : Block) : Block is
   return Block (X.O1 xor Y.O1, X.O2 xor Y.O2, X.O3 xor Y.O3, X.O4 xor Y.O4)
end function

function CYC (X : Block) : Block is
   return Block (
     Octet (X.O1.B2, X.O1.B3, X.O1.B4, X.O1.B5, X.O1.B6, X.O1.B7, X.O1.B8, X.O2.B1),
     Octet (X.O2.B2, X.O2.B3, X.O2.B4, X.O2.B5, X.O2.B6, X.O2.B7, X.O2.B8, X.O3.B1),
     Octet (X.O3.B2, X.O3.B3, X.O3.B4, X.O3.B5, X.O3.B6, X.O3.B7, X.O3.B8, X.O4.B1),
     Octet (X.O4.B2, X.O4.B3, X.O4.B4, X.O4.B5, X.O4.B6, X.O4.B7, X.O4.B8, X.O1.B1))
end function

function ADD (X, Y : Block) : Block is
   !implementedby "ADD" !external
   null -- sum modulo 2^32 of X and Y
end function

function CAR (X, Y : Block) : Block is
   !implementedby "CAR" !external
   null -- carry of the sum of X and Y; result is either x00000000 or x00000001
end function

function HIGH_MUL (X, Y : Block) : Block is
   !implementedby "HIGH_MUL" !external
   null -- 32 most significant bits of the 64-bit product of X and Y
end function

function LOW_MUL (X, Y : Block) : Block is
   !implementedby "LOW_MUL" !external
   null -- 32 least significant bits of the 64-bit product of X and Y
end function

end module

\end{lstlisting}

\subsection{The BLOCKVALUES library}

This library defines constant functions \T{x00000000}, ..., \T{xFFFFFFFF} that provide shorthand notations for block values. Only the useful constants (207 among $2^{32}$) are defined. An excerpt of this library is presented here.

\begin{lstlisting}
module BLOCKVALUES (OCTETVALUES, BLOCK) is

function x00000000 : Block is
  return Block (x00, x00, x00, x00)
end function

function x00000001 : Block is
  return Block (x00, x00, x00, x01)
end function
...
function xFFFFFFFE : Block is
   return Block (xFF, xFF, xFF, xFE)
end function

function xFFFFFFFF : Block is
   return Block (xFF, xFF, xFF, xFF)
end function

end module

\end{lstlisting}

\newpage

\subsection{The MAA specification}

\begin{lstlisting}
module MAA (BIT, OCTET, OCTETVALUES, BLOCK, BLOCKVALUES) is

!nat_bits 32

--------------------------------------------------------

type Message is
   list of Block
   with "==", "!=", "head", "tail", "append", "reverse"
end type

--------------------------------------------------------

function MakeMessage (N : Nat, in var INIT : Block, INCR : Block) : Message is
   assert N > 0;
   var I : Nat, RESULT : Message in
      RESULT := {};
      for I := 1 while I <= N by I := I + 1 loop
         RESULT := append (INIT, RESULT);
         INIT := ADD (INIT, INCR)
      end loop;
      return RESULT
   end var
end function

--------------------------------------------------------

function FIX1 (X : Block) : Block is
   return (X OR x02040801) AND xBFEF7FDF   -- A = x02040801, C = xBFEF7FDF
end function

--------------------------------------------------------

function FIX2 (X : Block) : Block is
   return (X OR x00804021) AND x7DFEFBFF   -- B = x00804021, D = x7DFEFBFF
end function

--------------------------------------------------------

function MUL1 (X, Y : Block) : Block is
   var U, L, S, C : Block in
      U := HIGH_MUL (X, Y);
      L := LOW_MUL (X, Y);
      S := ADD (U, L);
      C := CAR (U, L);
      assert (C == x00000000) or (C == x00000001);
      return ADD (S, C)
   end var
end function

--------------------------------------------------------

function MUL2 (X, Y : Block) : Block is
   var U, L, D, F, S, E, C : Block in
      U := HIGH_MUL (X, Y);
      L := LOW_MUL (X, Y);
      D := ADD (U, U);
      E := CAR (U, U);
      assert (E == x00000000) or (E == x00000001);
      F := ADD (D, ADD (E, E));
      S := ADD (F, L);
      C := CAR (F, L);
      assert (C == x00000000) or (C == x00000001);
      return ADD (S, ADD (C, C))
   end var
end function

--------------------------------------------------------

function MUL2A (X, Y : Block) : Block is
   var U, L, D, S, C : Block in
      U := HIGH_MUL (X, Y);
      L := LOW_MUL (X, Y);
      D := ADD (U, U);
      S := ADD (D, L);
      C := CAR (D, L);
      assert (C == x00000000) or (C == x00000001);
      return ADD (S, ADD (C, C))
   end var
end function

--------------------------------------------------------

function NeedAdjust (O : Octet) : Bool is
   return (O == x00) or (O == xFF)
end function

--------------------------------------------------------

function AdjustCode (O : Octet) : Bit is
   if NeedAdjust (O) then
      return 1
   else
      return 0
   end if
end function

--------------------------------------------------------

function Adjust (O, P : Octet) : Octet is
   if NeedAdjust (O) then
      return O xor P
   else
      return O
   end if
end function

--------------------------------------------------------

function PAT (X, Y : Block) : Octet is
   return Octet (AdjustCode (X.O1), AdjustCode (X.O2),
                 AdjustCode (X.O3), AdjustCode (X.O4),
                 AdjustCode (Y.O1), AdjustCode (Y.O2),
                 AdjustCode (Y.O3), AdjustCode (Y.O4))
end function

--------------------------------------------------------

function BYT (X, Y : Block, out U, L : Block) is
   var P : Octet in
      P := PAT (X, Y);
      U := Block (
         Adjust (X.O1, Octet (0, 0, 0, 0, 0, 0, 0, P.B1)),
         Adjust (X.O2, Octet (0, 0, 0, 0, 0, 0, P.B1, P.B2)),
         Adjust (X.O3, Octet (0, 0, 0, 0, 0, P.B1, P.B2, P.B3)),
         Adjust (X.O4, Octet (0, 0, 0, 0, P.B1, P.B2, P.B3, P.B4)));
      L := Block (
         Adjust (Y.O1, Octet (0, 0, 0, P.B1, P.B2, P.B3, P.B4, P.B5)),
         Adjust (Y.O2, Octet (0, 0, P.B1, P.B2, P.B3, P.B4, P.B5, P.B6)),
         Adjust (Y.O3, Octet (0, P.B1, P.B2, P.B3, P.B4, P.B5, P.B6, P.B7)),
         Adjust (Y.O4, Octet (P.B1, P.B2, P.B3, P.B4, P.B5, P.B6, P.B7, P.B8)))
   end var
end function

--------------------------------------------------------

function PreludeJ (J1 : Block,
                   out var J12, J14, J16 : Block, out J18 : Block,
                   out var J22, J24, J26 : Block, out J28 : Block) is
   J12 := MUL1 (J1, J1);
   J14 := MUL1 (J12, J12);
   J16 := MUL1 (J12, J14);
   J18 := MUL1 (J12, J16);
   J22 := MUL2 (J1, J1);
   J24 := MUL2 (J22, J22);
   J26 := MUL2 (J22, J24);
   J28 := MUL2 (J22, J26)
end function

--------------------------------------------------------

function PreludeK (K1 : Block,
                   out var K12, K14, K15, K17 : Block, out K19 : Block,
                   out var K22, K24, K25, K27 : Block, out K29 : Block) is
   K12 := MUL1 (K1, K1);
   K14 := MUL1 (K12, K12);
   K15 := MUL1 (K1, K14);
   K17 := MUL1 (K12, K15);
   K19 := MUL1 (K12, K17);
   K22 := MUL2 (K1, K1);
   K24 := MUL2 (K22, K22);
   K25 := MUL2 (K1, K24);
   K27 := MUL2 (K22, K25);
   K29 := MUL2 (K22, K27)
end function

--------------------------------------------------------

function Q (O : Octet) : Block is
   var B : Block in
      B := ADD (Block (x00, x00, x00, O), x00000001);
      return LOW_MUL (B, B)
   end var
end function

--------------------------------------------------------

function PreludeHJ (J14, J16, J18, J24, J26, J28 : Block,
                    out H4, H6, H8 : Block) is
   H4 := XOR (J14, J24);
   H6 := XOR (J16, J26);
   H8 := XOR (J18, J28)
end function

--------------------------------------------------------

function PreludeHK (K15, K17, K19, K25, K27, K29 : Block, P : Octet,
                    out var H0 : Block, out H5, H7, H9 : Block) is
   H0 := XOR (K15, K25);
   H5 := MUL2 (H0, Q (P));
   H7 := XOR (K17, K27);
   H9 := XOR (K19, K29)
end function

--------------------------------------------------------

function Prelude (in J, K : Block, out X, Y, V, W, S, T : Block) is
   var P : Octet,
       J1, J12, J14, J16, J18, J22, J24, J26, J28 : Block,
       K1, K12, K14, K15, K17, K22, K24, K25, K27, K19, K29 : Block,
       H4, H0, H5, H6, H7, H8, H9 : Block in
      BYT (J, K, ?J1, ?K1);
      P := PAT (J, K);
      PreludeJ (J1, ?J12, ?J14, ?J16, ?J18, ?J22, ?J24, ?J26, ?J28);
      use J12;
      use J22;
      PreludeK (K1, ?K12, ?K14, ?K15, ?K17, ?K19, ?K22, ?K24, ?K25, ?K27, ?K29);
      use K12;
      use K22;
      use K14;
      use K24;
      PreludeHJ (J14, J16, J18, J24, J26, J28, ?H4, ?H6, ?H8);
      PreludeHK (K15, K17, K19, K25, K27, K29, P, ?H0, ?H5, ?H7, ?H9);
      use H0;
      BYT (H4, H5, ?X, ?Y);
      BYT (H6, H7, ?V, ?W);
      BYT (H8, H9, ?S, ?T)
   end var
end function

--------------------------------------------------------

function MainLoop (in out X, Y, V : Block, W, B : Block) is
   V := CYC (V);
   var E, X1, Y1 : Block in
      E := XOR (V, W);
      X1 := MUL1  (XOR (X, B), FIX1 (ADD (XOR (Y, B), E)));
      Y1 := MUL2A (XOR (Y, B), FIX2 (ADD (XOR (X, B), E)));
      X := X1;
      Y := Y1
   end var
end function

--------------------------------------------------------

function Coda (in var X, Y, V : Block, W, S, T : Block, out Z : Block) is
   -- Coda (two more iterations with S and T)
   MainLoop (!?X, !?Y, !?V, W, S);
   MainLoop (!?X, !?Y, !?V, W, T);
   use V;
   Z := XOR (X, Y)
end function

--------------------------------------------------------

function MAC (J, K : Block, in var M : Message) : Block is
   !implementedby "MAC"
   -- this function is invoked externally from handwritten C code
   assert M != {};
   var X, X0, Y, Y0, V, V0, W, S, T, Z : Block, N : Nat in
      Prelude (J, K, ?X0, ?Y0, ?V0, ?W, ?S, ?T);
      X := X0;
      Y := Y0;
      V := V0;
      N := 0;
      loop
         MainLoop (!?X, !?Y, !?V, W, head (M));
         M := tail (M);
         N := N + 1;
         if M == {} then
            Coda (X, Y, V, W, S, T, ?Z);
            return Z
         elsif N == 256 then
            Coda (X, Y, V, W, S, T, ?Z);
            X := X0;
            Y := Y0;
            V := V0;
            N := 0;
            MainLoop (!?X, !?Y, !?V, W, Z)
         end if
      end loop
   end var
end function

--------------------------------------------------------

function CHECK : int is
   !implementedby "CHECK"
   -- this function is invoked externally from handwritten C code
   -- it checks the official test vectors given in [ISO 8730:1990] on the one
   -- hand, and [ISO 8731-2:1992] and [Davies-Clayden-88] on the other hand

   -- test vectors for function MUL1 - cf. Table 1 of [ISO 8731-2:1992]
   assert MUL1 (x0000000F, x0000000E) == x000000D2;
   assert MUL1 (xFFFFFFF0, x0000000E) == xFFFFFF2D;
   assert MUL1 (xFFFFFFF0, xFFFFFFF1) == x000000D2;

   -- test vectors for function MUL2 - cf. Table 1 of [ISO 8731-2:1992]
   assert MUL2 (x0000000F, x0000000E) == x000000D2;
   assert MUL2 (xFFFFFFF0, x0000000E) == xFFFFFF3A;
   assert MUL2 (xFFFFFFF0, xFFFFFFF1) == x000000B6;

   -- test vectors for function MUL2A - cf. Table 1 of [ISO 8731-2:1992]
   assert MUL2A (x0000000F, x0000000E) == x000000D2;
   assert MUL2A (xFFFFFFF0, x0000000E) == xFFFFFF3A;
   assert MUL2A (x7FFFFFF0, xFFFFFFF1) == x800000C2;
   assert MUL2A (xFFFFFFF0, x7FFFFFF1) == x000000C4;

   -- test vectors for function BYT - cf. Table 2 of [ISO 8731-2:1992]
   var U, L : Block in
      BYT (x00000000, x00000000, ?U, ?L);
      assert U == x0103070F;
      assert L == x1F3F7FFF;
      BYT (xFFFF00FF, xFFFFFFFF, ?U, ?L);
      assert U == xFEFC07F0;
      assert L == xE0C08000;
      BYT (xAB00FFCD, xFFEF0001, ?U, ?L);
      assert U == xAB01FCCD;
      assert L == xF2EF3501
   end var;

   -- test vectors for function PAT - cf. Table 2 of [ISO 8731-2:1992]
   assert PAT (x00000000, x00000000) == xFF;
   assert PAT (xFFFF00FF, xFFFFFFFF) == xFF;
   assert PAT (xAB00FFCD, xFFEF0001) == x6A;

   var J1, J12, J14, J16, J18, J22, J24, J26, J28 : Block,
       K1, K12, K14, K15, K17, K19, K22, K24, K25, K27, K29 : Block,
       H0, H4, H5, H6, H7, H8, H9 : Block, P : Octet in
      J1 := x00000100;
      K1 := x00000080;
      P  := x01;
      PreludeJ (J1, ?J12, ?J14, ?J16, ?J18, ?J22, ?J24, ?J26, ?J28);
      PreludeK (K1, ?K12, ?K14, ?K15, ?K17, ?K19, ?K22, ?K24, ?K25, ?K27, ?K29);
      PreludeHJ (J14, J16, J18, J24, J26, J28, ?H4, ?H6, ?H8);
      PreludeHK (K15, K17, K19, K25, K27, K29, P, ?H0, ?H5, ?H7, ?H9);

      -- test vectors for J1i values - cf. Table 3 of [ISO 8731-2:1992]
      assert J12 == x00010000;
      assert J14 == x00000001;
      assert J16 == x00010000;
      assert J18 == x00000001;

      -- test vectors for J2i values - cf. Table 3 of [ISO 8731-2:1992]
      assert J22 == x00010000;
      assert J24 == x00000002;
      assert J26 == x00020000;
      assert J28 == x00000004;

      -- test vectors for Hi values - cf. Table 3 of [ISO 8731-2:1992]
      assert H4 == x00000003;
      assert H6 == x00030000;
      assert H8 == x00000005;

      -- test vectors for K1i values - cf. Table 3 of [ISO 8731-2:1992]
      assert K12 == x00004000;
      assert K14 == x10000000;
      assert K15 == x00000008;
      assert K17 == x00020000;
      assert K19 == x80000000;

      -- test vectors for K2i values - cf. Table 3 of [ISO 8731-2:1992]
      assert K22 == x00004000;
      assert K24 == x10000000;
      assert K25 == x00000010;
      assert K27 == x00040000;
      assert K29 == x00000002;

      -- test vectors for Hi values - cf. Table 3 of [ISO 8731-2:1992]
      assert H0 == x00000018;
      assert Q (P) == x00000004;
      assert H5 == x00000060;
      assert H7 == x00060000;
      assert H9 == x80000002;

      -- test vectors for function PAT - cf. Table 3 of [ISO 8731-2:1992]
      assert PAT (H4, H5) == xEE;
      assert PAT (H6, H7) == xBB;
      assert PAT (H8, H9) == xE6;

      -- test vectors for function BYT - logically inferred from Table 3
      var U, L : Block in
         BYT (H4, H5, ?U, ?L);
         assert U == x01030703;
         assert L == x1D3B7760;
         BYT (H6, H7, ?U, ?L);
         assert U == x0103050B;
         assert L == x17065DBB;
         BYT (H8, H9, ?U, ?L);
         assert U == x01030705;
         assert L == x80397302
      end var
   end var;

   -- test vectors for function Main Loop - cf. Table 4 of [ISO 8731-2:1992]
   var A, B, C, D, E, F, G, M, V, W, X0, X, Y0, Y, Z : Block in
      -- first single-block message
      -- input values given in Table 4
      A  := x00000004;   -- fake "A" constant
      B  := x00000001;   -- fake "B" constant
      C  := xFFFFFFF7;   -- fake "C" constant
      D  := xFFFFFFFB;   -- fake "D" constant
      V  := x00000003;
      W  := x00000003;
      X0 := x00000002;
      Y0 := x00000003;
      M  := x00000005;
      -- loop iteration described page 10 of [ISO 8731-2:1992]
      V := CYC (V);            assert V == x00000006;
      E := XOR (V, W);         assert E == x00000005;
      X := XOR (X0, M);        assert X == x00000007;
      Y := XOR (Y0, M);        assert Y == x00000006;
      F := ADD (E, Y);         assert F == x0000000B;
      G := ADD (E, X);         assert G == x0000000C;
      F := OR (F, A);          assert F == x0000000F;
      G := OR (G, B);          assert G == x0000000D;
      F := AND (F, C);         assert F == x00000007;
      G := AND (G, D);         assert G == x00000009;
      X := MUL1 (X, F);        assert X == x00000031;
      Y := MUL2A (Y, G);       assert Y == x00000036;
      Z := XOR (X, Y);         assert Z == x00000007
   end var;

   var A, B, C, D, E, F, G, M, V, W, X0, X, Y0, Y, Z : Block in
      -- second single-block message
      -- input values given in Table 4
      A  := x00000001;   -- fake "A" constant
      B  := x00000004;   -- fake "B" constant
      C  := xFFFFFFF9;   -- fake "C" constant
      D  := xFFFFFFFC;   -- fake "D" constant
      V  := x00000003;
      W  := x00000003;
      X0 := xFFFFFFFD;
      Y0 := xFFFFFFFC;
      M  := x00000001;
      -- loop iteration described page 10 of [ISO 8731-2:1992]
      V := CYC (V);            assert V == x00000006;
      E := XOR (V, W);         assert E == x00000005;
      X := XOR (X0, M);        assert X == xFFFFFFFC;
      Y := XOR (Y0, M);        assert Y == xFFFFFFFD;
      F := ADD (E, Y);         assert F == x00000002;
      G := ADD (E, X);         assert G == x00000001;
      F := OR (F, A);          assert F == x00000003;
      G := OR (G, B);          assert G == x00000005;
      F := AND (F, C);         assert F == x00000001;
      G := AND (G, D);         assert G == x00000004;
      X := MUL1 (X, F);        assert X == xFFFFFFFC;
      Y := MUL2A (Y, G);       assert Y == xFFFFFFFA;
      Z := XOR (X, Y);         assert Z == x00000006
   end var;

   var A, B, C, D, E, F, G, M, V, W, X0, X, Y0, Y, Z : Block in
      -- third single-block message
      -- input values given in Table 4
      A  := x00000001;   -- fake "A" constant
      B  := x00000002;   -- fake "B" constant
      C  := xFFFFFFFE;   -- fake "C" constant
      D  := x7FFFFFFD;   -- fake "D" constant
      V  := x00000007;
      W  := x00000007;
      X0 := xFFFFFFFD;
      Y0 := xFFFFFFFC;
      M  := x00000008;
      -- loop iteration described page 10 of [ISO 8731-2:1992]
      V := CYC (V);            assert V == x0000000E;
      E := XOR (V, W);         assert E == x00000009;
      X := XOR (X0, M);        assert X == xFFFFFFF5;
      Y := XOR (Y0, M);        assert Y == xFFFFFFF4;
      F := ADD (E, Y);         assert F == xFFFFFFFD;
      G := ADD (E, X);         assert G == xFFFFFFFE;
      F := OR (F, A);          assert F == xFFFFFFFD;
      G := OR (G, B);          assert G == xFFFFFFFE;
      F := AND (F, C);         assert F == xFFFFFFFC;
      G := AND (G, D);         assert G == x7FFFFFFC;
      X := MUL1 (X, F);        assert X == x0000001E;
      Y := MUL2A (Y, G);       assert Y == x0000001E;
      Z := XOR (X, Y);         assert Z == x00000000
   end var;

   var A, B, C, D, E, F, G, M, V, W, X0, X, Y0, Y, Z : Block in
      -- three-block message: first block
      -- input values given in Table 4
      A  := x00000002;   -- fake "A" constant
      B  := x00000001;   -- fake "B" constant
      C  := xFFFFFFFB;   -- fake "C" constant
      D  := xFFFFFFFB;   -- fake "D" constant
      V  := x00000001;
      W  := x00000001;
      X0 := x00000001;
      Y0 := x00000002;
      M  := x00000000;
      -- loop iteration described page 10 of [ISO 8731-2:1992]
      V := CYC (V);            assert V == x00000002;
      E := XOR (V, W);         assert E == x00000003;
      X := XOR (X0, M);        assert X == x00000001;
      Y := XOR (Y0, M);        assert Y == x00000002;
      F := ADD (E, Y);         assert F == x00000005;
      G := ADD (E, X);         assert G == x00000004;
      F := OR (F, A);          assert F == x00000007;
      G := OR (G, B);          assert G == x00000005;
      F := AND (F, C);         assert F == x00000003;
      G := AND (G, D);         assert G == x00000001;
      X := MUL1 (X, F);        assert X == x00000003;
      Y := MUL2A (Y, G);       assert Y == x00000002;
      Z := XOR (X, Y);         assert Z == x00000001;

      -- three-block message: second block
      -- input values given in Table 4
      A  := x00000002;   -- fake "A" constant
      B  := x00000001;   -- fake "B" constant
      C  := xFFFFFFFB;   -- fake "C" constant
      D  := xFFFFFFFB;   -- fake "D" constant
      V  := x00000002;
      W  := x00000001;
      X0 := x00000003;
      Y0 := x00000002;
      M  := x00000001;
      -- loop iteration described page 10 of [ISO 8731-2:1992]
      V := CYC (V);            assert V == x00000004;
      E := XOR (V, W);         assert E == x00000005;
      X := XOR (X0, M);        assert X == x00000002;
      Y := XOR (Y0, M);        assert Y == x00000003;
      F := ADD (E, Y);         assert F == x00000008;
      G := ADD (E, X);         assert G == x00000007;
      F := OR (F, A);          assert F == x0000000A;
      G := OR (G, B);          assert G == x00000007;
      F := AND (F, C);         assert F == x0000000A;
      G := AND (G, D);         assert G == x00000003;
      X := MUL1 (X, F);        assert X == x00000014;
      Y := MUL2A (Y, G);       assert Y == x00000009;
      Z := XOR (X, Y);         assert Z == x0000001D;

      -- three-block message: third block
      -- input values given in Table 4
      A  := x00000002;   -- fake "A" constant
      B  := x00000001;   -- fake "B" constant
      C  := xFFFFFFFB;   -- fake "C" constant
      D  := xFFFFFFFB;   -- fake "D" constant
      V  := x00000004;
      W  := x00000001;
      X0 := x00000014;
      Y0 := x00000009;
      M  := x00000002;
      -- loop iteration described page 10 of [ISO 8731-2:1992]
      V := CYC (V);            assert V == x00000008;
      E := XOR (V, W);         assert E == x00000009;
      X := XOR (X0, M);        assert X == x00000016;
      Y := XOR (Y0, M);        assert Y == x0000000B;
      F := ADD (E, Y);         assert F == x00000014;
      G := ADD (E, X);         assert G == x0000001F;
      F := OR (F, A);          assert F == x00000016;
      G := OR (G, B);          assert G == x0000001F;
      F := AND (F, C);         assert F == x00000012;
      G := AND (G, D);         assert G == x0000001B;
      X := MUL1 (X, F);        assert X == x0000018C;
      Y := MUL2A (Y, G);       assert Y == x00000129;
      Z := XOR (X, Y);         assert Z == x000000A5
   end var;

   -- test vectors of Annex E.3.3 of [ISO 8730:1990]
   var A, B, C, D, E, F, G, M, V0, V, W, X0, X, Y0, Y : Block in
      A  := x02040801;   -- true "A" constant
      B  := x00804021;   -- true "B" constant
      C  := xBFEF7FDF;   -- true "C" constant
      D  := x7DFEFBFF;   -- true "D" constant
      X0 := x21D869BA;
      Y0 := x7792F9D4;
      V0 := xC4EB1AEB;
      W  := xF6A09667;
      M  := x0A202020;
      -- loop iteration on the first block M
      V := CYC (V0);           assert V == x89D635D7;
      E := XOR (V, W);         assert E == x7F76A3B0;
      X := XOR (X0, M);        assert X == x2BF8499A;
      Y := XOR (Y0, M);        assert Y == x7DB2D9F4;
      F := ADD (E, Y);         assert F == xFD297DA4;
      G := ADD (E, X);         assert G == xAB6EED4A;
      F := OR (F, A);          assert F == xFF2D7DA5;
      G := OR (G, B);          assert G == xABEEED6B;
      F := AND (F, C);         assert F == xBF2D7D85;
      G := AND (G, D);         assert G == x29EEE96B;
      X := MUL1 (X, F);        assert X == x0AD67E20;
      Y := MUL2A (Y, G);       assert Y == x30261492
   end var;

   -- test vectors for the whole algorithm - cf. Table 5 of [ISO 8731-2:1992]
   var J, K, X, Y, V, W, S, T, Z, M1, M2 : Block in
      -- first column of Table 5
      J  := x00FF00FF;
      K  := x00000000;
      M1 := x55555555;
      M2 := xAAAAAAAA;
      assert PAT (J, K) == xFF;
      Prelude (J, K, ?X, ?Y, ?V, ?W, ?S, ?T);
      assert X == x4A645A01;
      assert Y == x50DEC930;
      assert V == x5CCA3239;
      assert W == xFECCAA6E;
      assert S == x51EDE9C7;
      assert T == x24B66FB5;
      -- 1st MainLoop iteration
      MainLoop (!?X, !?Y, !?V, W, M1);
      assert X == x48B204D6;
      assert Y == x5834A585;
      -- 2nd MainLoop iteration
      MainLoop (!?X, !?Y, !?V, W, M2);
      assert X == x4F998E01;
      assert Y == xBE9F0917;
      -- Coda: MainLoop iteration with S
      MainLoop (!?X, !?Y, !?V, W, S);
      assert X == x344925FC;
      assert Y == xDB9102B0;
      -- Coda: MainLoop iteration with T
      MainLoop (!?X, !?Y, !?V, W, T);
      use V;
      assert X == x277B4B25;
      assert Y == xD636250D;
      Z  := XOR (X,Y);
      assert Z == xF14D6E28
   end var;

   var J, K, X, Y, V, W, S, T, Z, M1, M2 : Block in
      -- second column of Table 5
      J  := x00FF00FF;
      K  := x00000000;
      M1 := xAAAAAAAA;
      M2 := x55555555;
      assert PAT (J, K) == xFF;
      Prelude (J, K, ?X, ?Y, ?V, ?W, ?S, ?T);
      assert X == x4A645A01;
      assert Y == x50DEC930;
      assert V == x5CCA3239;
      assert W == xFECCAA6E;
      assert S == x51EDE9C7;
      assert T == x24B66FB5;
      -- 1st MainLoop iteration
      MainLoop (!?X, !?Y, !?V, W, M1);
      assert X == x6AEBACF8;
      assert Y == x9DB15CF6;
      -- 2nd MainLoop iteration
      MainLoop (!?X, !?Y, !?V, W, M2);
      assert X == x270EEDAF;
      assert Y == xB8142629;
      -- Coda: MainLoop iteration with S
      MainLoop (!?X, !?Y, !?V, W, S);
      assert X == x29907CD8;
      assert Y == xBA92DB12;
      -- Coda: MainLoop iteration with T
      MainLoop (!?X, !?Y, !?V, W, T);
      use V;
      assert X == x28EAD8B3;
      assert Y == x81D10CA3;
      Z  := XOR (X,Y);
      assert Z == xA93BD410
   end var;

   var J, K, X, Y, V, W, S, T, Z, M1, M2 : Block in
      -- third column of Table 5
      J  := x55555555;
      K  := x5A35D667;
      M1 := x00000000;
      M2 := xFFFFFFFF;
      assert PAT (J, K) == x00;
      Prelude (J, K, ?X, ?Y, ?V, ?W, ?S, ?T);
      assert X == x34ACF886;
      assert Y == x7397C9AE;
      assert V == x7201F4DC;
      assert W == x2829040B;
      assert S == x9E2E7B36;
      assert T == x13647149;
      -- 1st MainLoop iteration
      MainLoop (!?X, !?Y, !?V, W, M1);
      assert X == x2FD76FFB;
      assert Y == x550D91CE;
      -- 2nd MainLoop iteration
      MainLoop (!?X, !?Y, !?V, W, M2);
      assert X == xA70FC148;
      assert Y == x1D10D8D3;
      -- Coda: MainLoop iteration with S
      MainLoop (!?X, !?Y, !?V, W, S);
      assert X == xB1CC1CC5;
      assert Y == x29C1485F;
      -- Coda: MainLoop iteration with T
      MainLoop (!?X, !?Y, !?V, W, T);
      use V;
      assert X == x288FC786;
      assert Y == x9115A558;
      Z  := XOR (X,Y);
      assert Z == xB99A62DE
   end var;

   var J, K, X, Y, V, W, S, T, Z, M1, M2 : Block in
      -- fourth column of Table 5
      J  := x55555555;
      K  := x5A35D667;
      M1 := xFFFFFFFF;
      M2 := x00000000;
      assert PAT (J, K) == x00;
      Prelude (J, K, ?X, ?Y, ?V, ?W, ?S, ?T);
      assert X == x34ACF886;
      assert Y == x7397C9AE;
      assert V == x7201F4DC;
      assert W == x2829040B;
      assert S == x9E2E7B36;
      assert T == x13647149;
      -- 1st MainLoop iteration
      MainLoop (!?X, !?Y, !?V, W, M1);
      assert X == x8DC8BBDE;
      assert Y == xFE4E5BDD;
      -- 2nd MainLoop iteration
      MainLoop (!?X, !?Y, !?V, W, M2);
      assert X == xCBC865BA;
      assert Y == x0297AF6F;
      -- Coda: MainLoop iteration with S
      MainLoop (!?X, !?Y, !?V, W, S);
      assert X == x3CF3A7D2;
      assert Y == x160EE9B5;
      -- Coda: MainLoop iteration with T
      MainLoop (!?X, !?Y, !?V, W, T);
      use V;
      assert X == xD0482465;
      assert Y == x7050EC5E;
      Z  := XOR (X,Y);
      assert Z == xA018C83B
   end var;

   var J, K, X, Y, V, W, S, T : Block in
      -- test vectors of Annex E.3.3 of [ISO 8730:1990]
      J  := xE6A12F07;
      K  := x9D15C437;
      Prelude (J, K, ?X, ?Y, ?V, ?W, ?S, ?T);
      assert X == x21D869BA;
      assert Y == x7792F9D4;
      assert V == xC4EB1AEB;
      assert W == xF6A09667;
      assert S == x6D67E884;
      assert T == xA511987A
   end var;

   -- test vectors for the whole algorithm
   var B, J, K, X, Y, V, W, S, T : Block, M : Message in
      J := x80018001;
      K := x80018000;

      -- test mentioned in Table 6 of [ISO 8731-2:1992]
      -- iterations on a message containg 20 null blocks
      Prelude (J, K, ?X, ?Y, ?V, ?W, ?S, ?T);
      B := x00000000;
      -- 1st MainLoop iteration
      MainLoop (!?X, !?Y, !?V, W, B);
      assert X == x303FF4AA;
      assert Y == x1277A6D4;
      -- 2nd MainLoop iteration
      MainLoop (!?X, !?Y, !?V, W, B);
      assert X == x55DD063F;
      assert Y == x4C49AAE0;
      -- 3rd MainLoop iteration
      MainLoop (!?X, !?Y, !?V, W, B);
      assert X == x51AF3C1D;
      assert Y == x5BC02502;
      -- 4th MainLoop iteration
      MainLoop (!?X, !?Y, !?V, W, B);
      assert X == xA44AAAC0;
      assert Y == x63C70DBA;
      -- 5th MainLoop iteration
      MainLoop (!?X, !?Y, !?V, W, B);
      assert X == x4D53901A;
      assert Y == x2E80AC30;
      -- 6th MainLoop iteration
      MainLoop (!?X, !?Y, !?V, W, B);
      assert X == x5F38EEF1;
      assert Y == x2A6091AE;
      -- 7th MainLoop iteration
      MainLoop (!?X, !?Y, !?V, W, B);
      assert X == xF0239DD5;
      assert Y == x3DD81AC6;
      -- 8th MainLoop iteration
      MainLoop (!?X, !?Y, !?V, W, B);
      assert X == xEB35B97F;
      assert Y == x9372CDC6;
      -- 9th MainLoop iteration
      MainLoop (!?X, !?Y, !?V, W, B);
      assert X == x4DA124A1;
      assert Y == xC6B1317E;
      -- 10th MainLoop iteration
      MainLoop (!?X, !?Y, !?V, W, B);
      assert X == x7F839576;
      assert Y == x74B39176;
      -- 11th MainLoop iteration
      MainLoop (!?X, !?Y, !?V, W, B);
      assert X == x11A9D254;
      assert Y == xD78634BC;
      -- 12th MainLoop iteration
      MainLoop (!?X, !?Y, !?V, W, B);
      assert X == xD8804CA5;
      assert Y == xFDC1A8BA;
      -- 13th MainLoop iteration
      MainLoop (!?X, !?Y, !?V, W, B);
      assert X == x3F6F7248;
      assert Y == x11AC46B8;
      -- 14th MainLoop iteration
      MainLoop (!?X, !?Y, !?V, W, B);
      assert X == xACBC13DD;
      assert Y == x33D5A466;
      -- 15th MainLoop iteration
      MainLoop (!?X, !?Y, !?V, W, B);
      assert X == x4CE933E1;
      assert Y == xC21A1846;
      -- 16th MainLoop iteration
      MainLoop (!?X, !?Y, !?V, W, B);
      assert X == xC1ED90DD;
      assert Y == xCD959B46;
      -- 17th MainLoop iteration
      MainLoop (!?X, !?Y, !?V, W, B);
      assert X == x3CD54DEB;
      assert Y == x613F8E2A;
      -- 18th MainLoop iteration
      MainLoop (!?X, !?Y, !?V, W, B);
      assert X == xBBA57835;
      assert Y == x07C72EAA;
      -- 19th MainLoop iteration
      MainLoop (!?X, !?Y, !?V, W, B);
      assert X == xD7843FDC;
      assert Y == x6AD6E8A4;
      -- 20th MainLoop iteration
      MainLoop (!?X, !?Y, !?V, W, B);
      assert X == x5EBA06C2;
      assert Y == x91896CFA;
      -- Coda: MainLoop iteration with S
      MainLoop (!?X, !?Y, !?V, W, S);
      assert X == x1D9C9655;
      assert Y == x98D1CC75;
      -- Coda: MainLoop iteration with T
      MainLoop (!?X, !?Y, !?V, W, T);
      use V;
      assert X == x7BC180AB;
      assert Y == xA0B87B77;
      M := MakeMessage (20, x00000000, x00000000);
      assert MAC (J, K, M) == xDB79FBDC;

      -- supplementary tests added by H. Garavel and L. Marsso
      M := MakeMessage (16, x00000000, x07050301);
      assert MAC (J, K, M) == x8CE37709;

      M := MakeMessage (256, x00000000, x07050301);
      assert MAC (J, K, M) == x717153D5;

      M := MakeMessage (4100, x00000000, x07050301);
      assert MAC (J, K, M) == x7783C51D
   end var;
   return 0
end function

end module

\end{lstlisting}

\end{document}